# ToffA-DSPL: An approach of Trade-off Analysis for designing Dynamic Software Product Lines

MICHELLE LARISSA LUCIANO CARVALHO*, and EDUARDO SANTANA DE ALMEIDA, Federal University of Bahia (UFBA), Computer Institute, Brazil

ISMAYLE DE SOUSA SANTOS*, State University of Ceará (UECE), Computer Institute, Brazil

PAULO CESAR MASIERO, University of Sao Paulo (USP), Computer Science Department, Brazil

Software engineers have adopted the Dynamic Software Product Lines (DSPL) engineering practices to develop Dynamically Adaptable Software (DAS). DAS is seen as a DSPL application and must cope with a large number of configurations of features, Non-functional Requirements (NFRs), and contexts. However, the accurate representation of the impact of features over NFRs and contexts for the identification of optimal configurations is not a trivial task. Software engineers need to have domain knowledge and design DAS before deploying to satisfy those requirements. Aiming to handle them, we proposed an approach of Trade-off Analysis for DSPL at design-time, named ToffA-DSPL. It deals with the configuration selection process considering interactions between NFRs and contexts. We performed an exploratory study based on simulations to identify the usefulness of the ToffA-DSPL approach. In general, the configurations suggested by ToffA-DSPL provide high satisfaction levels of NFRs. Based on simulations, we evidenced that our approach aims to explore reuse and is useful for generating valid and optimal configurations. In addition, ToffA-DSPL enables software engineers to conduct trade-off analysis, evaluate changes in the context feature, and define an adaptation model from optimal configurations found in the analysis.

# 1 INTRODUCTION

Dynamic Software Product Lines (DSPL) engineering is a paradigm aimed at providing software systems with the capability of handling adaptations at runtime. A DSPL application monitors the environment and adapts its behavior according to changes in the execution environment or user requests. Then, software engineers used

---

*Corresponding author.

Authors' addresses: Michelle Larissa Luciano Carvalho, michellellc@ufba.br; Eduardo Santana de Almeida, esa@dcc.ufba.br, Federal University of Bahia (UFBA), Computer Institute, Salvador, Bahia, Brazil; Ismayle de Sousa Santos, State University of Ceará (UECE), Computer Institute, Ceará, Brazil, ismayle.santos@uece.br; Paulo Cesar Masiero, University of Sao Paulo (USP), Computer Science Department, São Paulo, Brazil, masiero@icmc.usp.br.

DSPL engineering to provide support for developing Dynamically Adaptable Software (DAS) by capturing and managing dynamic variability. A DAS itself is seen as a DSPL application [8].

Dynamic variability occurs due to product variations following the contextual changes [7]. Thus, it consists of product customization that can happen at runtime. This customization to context changes corresponds to product derivation in the product line terminology [35]. In this scenario, DSPL applications should be capable of handling context identification and must be prepared to dynamically adapt their behavior to meet distinct scenarios. Such adaptations comprise activating and deactivating software system features [7].

Additionally, DAS is a software system capable of adapting at runtime based on changes in the surrounding environment. Aiming to design and develop this kind of system, the software engineer must satisfy: the **system's features** – *"aspects valuable to the customer..."* [62] *or concerning an attribute of a system that directly affects end-users* [44]. A formalized definition of system's features is given in [10]: *"a logical unit of behavior that is specified by a set of functional and quality requirements"*; **contexts** – *information computationally accessible and upon which behavioral variations depend* [14]; and **Non-functional Requirements (NFRs)** – *internal system properties and are recognized as an important factor for the success of software projects* [73]. However, it is not a trivial task to satisfy such elements at the same time. In this sense, practitioners have used DSPL engineering processes to support DAS development. This has helped DAS developers to cope with context variability and a large number of configurations.

## 1.1 Motivation

According to Bencomo *et. al.* [6], DSPL applications should be prepared to deal with the following dimensions of variability: *structural variability* and *environment* or *context variability*. These dimensions can be modeled as *variation points*, *i.e.*, as specific locations where decisions can be made to express *variants* (product configurations) [36]. In the first dimension, the variation points represent different configurations that can be derived by considering the constraints defined in the model. Thus, it refers to the configuration of the system's features. In the second dimension, the variation points represent the properties of the environment. It represents the variants of contextual information relevant to the system based on the environment where it resides [12, 53].

Although DSPL product derivation occurs at runtime, the software engineer must identify at design time the possible adaptations and the information that can affect the product configuration. This encompasses *variability modeling*, which consists of one of the most important activities in DSPL engineering [12]. It aids software engineers in handling the system's features, contexts, and NFRs. Sousa *et.al.* [17] reported a set of research opportunities related to quality evaluations of DSPL, such as the definition of thresholds for quality measures, the development of approaches for prioritization of NFRs according to DSPL operations and domains, besides the conflict mapping among NFRs. The authors state that the system quality evaluation is not a trivial task and must be made not only at runtime but also at design time to check the system's capacity to meet self-adaptive operations.

## 1.2 Problem

Dealing with NFRs makes the configuration selection process more difficult since such properties tend to be qualitative and may not be easily mathematically quantifiable (*e.g.*, specifying resiliency and efficiency) [77, 80]. Quantifying NFRs often relies on domain knowledge and may not be optimal in a specific DSPL, which must meet certain contextual changes [80]. Indeed, among the major challenges that software engineers face with regard to the modeling and development of DSPL applications [12, 72, 74] the following are highlighted: *(i)* checking the existing or new constraints of the feature model to avoid unfeasible products [5] and the DSPL behavior that can take different configurations due to changes in context [43]; *(ii)* the accurate representation of the impact of features over contexts and NFRs aiming to identify optimal configurations that satisfy the stakeholder's preferences and constraints [72]; and *(iii)* prediction of a set of possible dynamic adaptations that may occur following several changes in the environment behavior [12, 74];

Firstly, it is necessary to address since design-time the dimensions of *structural variability* and *context variability* since unanticipated conditions can trigger failures and inconsistent reconfigurations at runtime. To avoid such failures and inconsistencies, software engineers can simulate at design time, a set of possible adaptations that meet the variability dimensions. Next, s/he can choose which adaptations should be developed based on those that satisfy specific quality requirements. These dimensions are handled during both, *domain engineering* and *application engineering* processes aiming to model and develop DSPL by exploring adaptability [34]. Indeed, DSPL engineering processes aim to design applications through the generation of a huge number of configurations and to help make software systems extensible, besides achieving a good quality of service. However, those numerous configurable options result in a software configuration space explosion and lead to real challenges for developers. This explosion makes the analysis more difficult, as different configurations can be conflated together and generally complicates the application understanding tasks [61].

Secondly, software engineers must find a system's feature combination that can simultaneously satisfy constraints specified in feature and context models, NFRs, and stakeholder preferences. It means that they have to measure many configurations until they find the optimal ones. Thus, the product configuration process in DSPL engineering can be viewed as a complex optimization problem [72]. When dealing with feature selection to meet desired quality objectives in DSPL, most of the existing studies are not focused on the interactions between contextual information and NFRs. In addition, such studies do not use any strategy to support the selection of the most suitable configuration [32]. As pointed out by Huebscher *et.al.* [41], there is a planning type named *Utility-based* planning that enables us to find optimal configurations by the utility values that represent the desirable variants. Such utility value is approximated by a utility function over contexts and NFRs. Therefore, the optimal configurations can be found using a heuristic that approximates the impact of features over contexts and NFRs in the utility value.

To the best of our knowledge, few studies [3, 23, 27, 30, 31, 35, 55, 57, 58, 69, 71] propose an approach to identify optimally (or nearly optimal) configurations by satisfying the interactions between *contexts* and *NFRs*, *i.e.*, these interactions are left aside in the variability modeling task. In addition, many of the existing studies addressing this issue do not use any strategy for dealing with configuration selection conflicts [32]. Studies presented by Hallsteinsen *et. al.* [35], Franco *et. al.* [25], Edwards *et. al.* [21], Paucar *et. al.* [59], Esfahani *et. al.* [23], Greenwood *et. al.* [30], Guedes *et. al.* [31], Nascimento *et. al.* [55], and Sanchez *et. al.* [69] use the utility function to approximate the fulfillment of stakeholder's preferences in different situations. Among them, only Franco *et. al.* [25], Edwards *et. al.* [21], and Pacuar *et. al.* [59] do not consider the trade-off between contextual information and NFRs in the decision-making. Additionally, Hallsteinsen *et. al.* [35], Esfahani *et. al.* [23], Greenwood *et. al.* [30], Guedes *et. al.* [31], Nascimento *et. al.* [55], and Sanchez *et. al.* [69] propose approaches that model the interactions between contexts and NFRs. Although those approaches assume the *Utility-based* planning as a strategy to formalize the knowledge obtained and deal with the interactions between such conflicting elements, the authors do not consider or apply all modeling characteristics (*prioritization*, *satisfaction levels*, and *contribution*) to system's features, context, and NFRs (see section 3). Indeed, this is an important research gap, since the number of product configurations increases exponentially with the number of features and many configurations satisfy the same requirements.

In our previous work, we investigated DSPL verification, context variability modeling, and issues related to the implementation of DSPL. Hence, we have proposed a DSPL model checking technique [70] and feature model language for DSPL that may specify constraints among context [18]. Also, we evidenced in our previous study [13] that DSPL should also deal with evolution in terms of functionality and adaptation capabilities when the demand arises for new requirements for existing products or new configurations. However, changes and extensions to the design may affect problem and solution spaces. The first refers to the system's specifications established during the domain analysis, whereas the second refers to assets created during the design and implementation phases [9]. In general, evolving a variability model may affect the solution space and vice versa, *i.e.*, the evolution can lead to inconsistencies within a given space and affect qualities that DSPL must have.

## 1.3 Overview and contributions

Based on the aforementioned problem, we decided to investigate DSPL engineering from a variability-modeling perspective. We are pursuing a threefold goal in DSPL engineering field by proposing an approach that *(i)* manages **at design-time** both dimensions, *structural variability* and *context variability*; *(ii)* facilitates the understanding of how DSPL applications can behave from a certain context change, and *(iii)* enables software engineers to conduct trade-off analysis to find optimal configurations that meet the constraints and the interactions between contexts and NFRs. Our work is related to the *reactive* approach [46], which promotes software mass customization, since a minimum number of products must be incorporated in advance. In addition, it can be handy when software engineers need assistance in understanding how to design a variety of configurable options for DSPL applications. Such an approach is based on the principle that each configuration option must be optimal to meet certain contextual changes without losing service quality. Indeed, our approach aims to explore reuse embracing domain analysis, modeling, checking, prioritization, contribution, and optimization to generate valid and optimal configurations.

Additionally, this work reviews and extends our previous work [13, 18, 70] to provide a comprehensive approach to support the trade-off analysis of DAS at design-time, named ToffA-DSPL. It deals with the configuration selection process embracing interactions between contextual information and NFRs. Additionally, we defined an optimization model that recommends optimal configurations by considering all integrity constraints and variability of DSPL models. In this sense, we used *utility-based* planning [12] as a strategy to deal with trade-off analysis. The contributions of this work are described as follows:

- We defined the ToffA-DAS approach to deal with the configuration selection process embracing interactions between contextual information and NFRs. Such an approach uses a feature model with constraints among contexts. In addition, ToffA-DAS is based on a utility function that makes it possible to define a weighted mean of the differences between properties describing the service provided by the application and properties representing user needs. The weights, *i.e.* the utility values represent stakeholder's priorities [35]. Based on the utility values, we define an optimization model that recommends optimal configurations by considering a diversity of adaptation rules and constraints of the feature model;
- We proposed a technique to conduct trade-off analysis in DSPL engineering by changing the prioritization of modeling elements. This is important due to interactions among adaptations triggered by different contexts activated at the same time. Indeed, such a trade-off analysis should be done during the requirements specification, and decisions should be taken by the software engineers aiming to satisfy the needs of their stakeholders; and
- We also proposed a strategy to analyze context changes in order to define adaptation models. Thus, software engineers can define such adaptation models for each prioritization of contexts and NFRs. The advantage of this model is to support predicting a set of possible adaptations at design time, then it can be eventually developed and handled at runtime.

## 1.4 Organization of the paper

The remainder of this paper is organized as follows. Section 2 presents the background in the area. Section 3 discusses related work. The approach overview is described in Section 4. In addition, it describes the optimization model defined in our approach. In Section 5, we describe how our approach can support software engineers for the reasoning of adaptability. Section 6 reports the lessons learned. The limitations of the approach are presented in Section 7. Section 8 presents concluding remarks and points out future directions.

## 2 BACKGROUND

This section provides background concepts on the topics involved in this investigation, namely Dynamic Software Product Lines (DSPL), Extended Context-aware Feature Modeling (eCFM), requirements engineering, planning the development of DSPL projects, and variability modeling.

### 2.1 Dynamic Software Product Lines (DSPL)

Highly configurable systems increase the need to deal with variability at runtime because they modify their internal structures dynamically and, consequently, their behavior in response to internal and external incentives [51]. For this reason, some researchers introduced the DSPL approach to deal with changes in the environment and user requests during runtime.

The development of DSPL applications involves two essential activities: *monitoring* the current situation for detecting events that might require adaptation and *controlling* the adaptation through the management of variability. In that case, it is important to analyze the change's impact on the product's requirements or constraints and plan for deriving a suitable adaptation to cope with new situations. In addition, these activities encompass some properties, such as *automatic decision-making, autonomy and adaptivity*, and *context-awareness* [7, 34].

The DSPL engineering extends the Software Product Lines (SPL) engineering principles to deal with the modeling and implementation of DAS [18, 34]. A product derived from DSPL differs from other instantiated SPL by the capacity to adapt through the binding of variation points at runtime. A modification when detected in the operational environment activates the product reconfiguration to provide context-relevant services or collect quality requests (including safety, reliability, and performance) [47]. In this scenario, any flaw in the reconfiguration of the DSPL application directly impacts the user experience, because it happens when the system is already under its control [16]. Therefore, it is necessary to investigate the aspects of the reconfiguration of product lines at design time aiming to deliver high-quality software.

### 2.2 Requirements Engineering

NFRs are considered constraints imposed on software or qualities that the software product must have. According to these two perspectives, NFRs can be identified by considering several elements such as constraints, concerns, goals, and quality attributes. Their relevance degree can also vary depending on the different software types or application domains. Though NFRs are as important as functional requirements, they are neglected, poorly understood, and not adequately considered in the development of single systems and traditional product lines [50]. The same happens in the DSPL engineering field since just a few studies address product configuration issues considering NFRs information [17, 73]. Poor requirements elicitation results in many failures in software systems [16]. For this reason, it is essential to understand how a system must behave aiming for its dependability.

Goal-oriented Requirement Engineering (GORE) provides a means to support the requirements elicitation and decompose them into well-defined entities and reason about the alternatives to meet them [52]. In this way, goal-oriented modeling has been recognized as a suitable strategy to represent the objectives of a software system and stakeholders' preferences [40]. Goal models may then be used as part of the entire system life-cycle, in particular, to improve the continuous delivery of software systems [1]. Goal models fit well with the early domain engineering phases by supporting different alternatives for satisfying stakeholder's preferences [2, 81]. They are composed of *goals*, *hard goals*, and *soft goals* depending on the precision of their satisfaction level in terms of stakeholder's intentions. *Goals* consist of the objectives of the stakeholders. *Hard goals* represent the system's features and have satisfaction level defined clearly, whereas *soft goals* express NFRs and have a subjective satisfaction level that cannot be defined in a clear-cut way and cannot be fully evaluated [29]. Thus, *hard goals* and *soft goals* may be judged as satisfied or unsatisfied to different degrees and at different stages of the development of DSPL applications [2].

The GORE approach supports the conceptual modeling of variable and common requirements of a DSPL application by representing them as goals [2]. Such *goals* consist of objectives or desirable states for software systems and are decomposed through AND/OR relationships with *hard goals*. In the AND relationship, all *hard goals* should satisfy its parent *goal*, whereas, in the OR relationship, at least one *hard goal* satisfies its parent *goal* [81] (see Figure 4).

## 2.3 Planning the development of DSPL projects

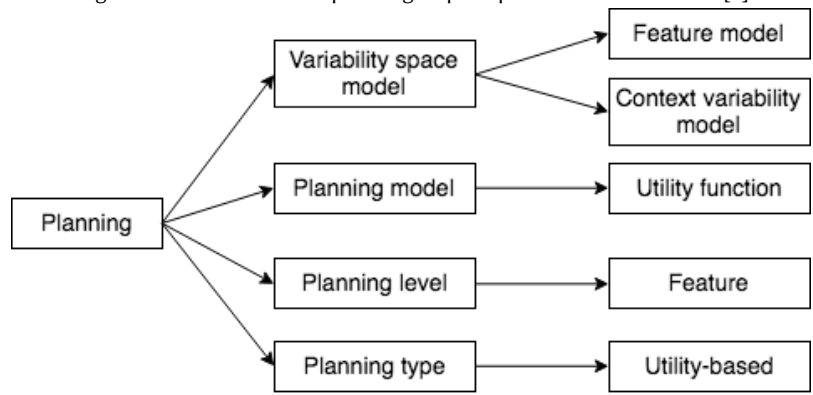

Fig. 1. Dimensions related planning step adapted from Bashari *et al.* [4]

Bashari et al. [4] proposed a framework to classify different dimensions of adaptation realization in DSPL projects. The framework dimensions are organized in a taxonomy that uses the steps of the MAPE-K loop. Hinchey et al. [49] proposed the development of DSPL applications based on the MAPE-K loop. It defines how systems adapt their behavior to keep their goals controlled based on control systems or optimization strategies. The MAPE-K loop is composed of four activities: Monitor, Analyze, Plan, and Execute, including the base of Knowledge to support model representations. Thus, it supports the autonomic condition of DSPL by mapping contexts to variants and offering them a better selection of the possible alternatives according to goals to be accomplished [12].

In this work, we focus on the planning step (see Figure 1) to define a *planning model* based on the *utility function*. It aims to support the software engineers in decision-making during the modeling initial phase of DSPL projects. The DSPL community has devoted most efforts to developing planning approaches [7]. Hallsteinsen *et. al.* [35], for instance, defined a *utility-based* approach to representing the desirability of the current configuration in the current context. Thus, the goal of adaptation is the maximization of the value of this function by considering the system properties and its context. In cases when the utility value is unacceptable, the planner uses a brute-force technique to find a configuration with the highest predicted utility value and adapts to it [4].

In general, a planner uses information provided in the context model to decide whether an adaptation is necessary and find, within the *variability space model*, the variant that satisfies such an adaptation [4]. Each desired variant is represented by a system's feature that, in turn, is specified as a *planning level*. Therefore, the planner decides what features should be active in the system according to its context.

The *planning type* is related to the strategies that are employed for selecting the most suitable variant of the system. According to Huebscher *et. al.* [41], the *planning types* can be categorized into three groups, as follows:

- *Rule-based* planning: This planning strategy employs ECA rules (Event Condition-Action) or state-transition diagrams to define adaptations at design time. Some *rule-based* planners allow the modification of the rules at runtime [54]. However, the usage of this planning requires not only the enumeration of possible system reconfigurations but also the thorough knowledge of operating the environment at design time. Indeed, *rule-based* planning requires that the software engineer have in-depth expertise of the operating environment [7].
- *Goal-based* planning: Using this planning, the possible adaptations that can occur to meet contextual changes are figured out at runtime, unlike *rule-based* planning where such actions are specified at design time. Therefore, the high-level goals of the adaptation are formally defined and figured out by the planner. Afterward, the problem of detecting the most suitable action is reduced to a satisfiability problem (SAT) [56, 64] or a constraint satisfaction problem (CSP) [20, 57, 71], which is then solved using an SAT-solver or a CSP-solver. In this way, the adaptation goals are expressed by constraints *over* system quality; and
- *Utility-based* planning: In this planning strategy, the problem of feature selection may be represented as mono-objective or multi-objective optimization in order to find the optimal configuration for that feature model [58, 65]. For this, it is necessary to predict the utility function(s) to approximate the fulfillment of stakeholders' preferences in different situations. This prediction function is used to find the configuration that has the highest predicted utility value. Such utility consists of a quantitative value to represent the weight of the system's feature [28].

We investigated the *state-of-the-art* in the search of a planning strategy that enables to support *(i)* on prediction of a set of possible adaptations following the changes in the environment-behavior and *(ii)* in the quantitative representation of the impact of features over contexts and NFRs aiming to identify optimal configurations, which satisfy the stakeholder's preferences and constraints. As a result, we noticed that the use of the *utility-based* planning enables us to find optimal configurations following the quantitative values (utility values) that represent the desirable variants. Such utility value is approximated by a utility function over contexts and NFRs. Therefore, the optimal configurations can be found using a heuristic that approximates the impact of features over these elements in the utility value.

## 2.4 Variability Modeling

In DSPL engineering, feature models are commonly used as a variability model due to their power to represent the complex variability of software systems. Nevertheless, a context variability model also must be defined in this *variability space modeling* since it represents the contextual triggers for variation [4]. Context variability deals with the diversity of the contextual changes that influence the dynamic behavior of systems [12]. Hartmann *et al.* [36] introduced the notion of context variability to identify context features and capture the common and variable information of contexts during the initial modeling phase. In DSPL engineering, the variability modeling task handles both the system's features and context features as well as dependencies between them. The software engineers, in turn, use the feature model to manage system variability. It consists of a model language widely employed to represent the system's features in a hierarchical structure [44]. However, such language was extended with the purpose also to modeling the context variability.

An expressive and correct model for DAS development is a first-class concern. In this section, we present the steps for modeling DAS using the extended Context-aware Feature Modeling (eCFM) technique, which was proposed in our previous study [18]. The eCFM technique consists of an extension of the Context-Aware Feature Model [67] to deal with constraints among contexts. Figure 2 shows a running example in the self-adaptive wireless sensor network domain, named *GridStix*, which was adapted from Sawyer *et al.* [71]. To use the eCFM, the domain/software engineer must perform the *context analysis* task [12] during the DSPL domain analysis, identifying context-aware properties and how they affect the system configuration. A literature review and an analysis of existing applications can support this task.

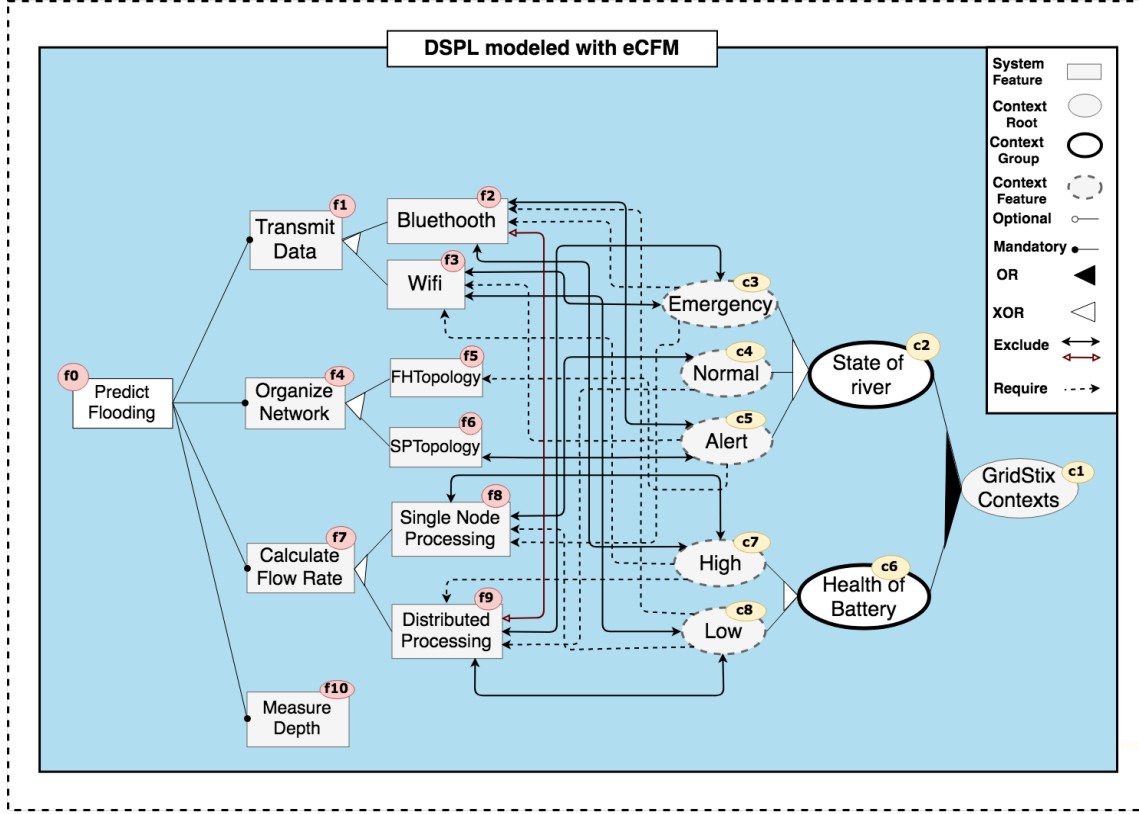

Fig. 2. GridStix DAS modeled with eCFM. It represents all features, context features, context groups, context root, and their respective relationships.

In particular, stakeholders should identify four main elements for specifying DSPL in eCFM: *(i)* **functional requirements** represented as system features; *(ii)* **variability information** specified as feature relationships of the type mandatory, optional, OR-group or XOR-group; *(iii)* **context information**, which are described as context states; and *(iv)* **adaptations rules**, which are specified as dependencies among features and contexts.

The main concepts used in eCFM are represented, as follows:

- **Context Feature** – a relevant context state to the software, which can be active or inactive, *i.e.*, the context can be active or not in the surrounding environment;
- **Context Group** – a set of context features that can specify XOR-group/OR-group relationships or optional context features. For example, the XOR-the relationship between **High** and **Low** in Figure 2.
- **Context Root** – a more abstract context model that aggregates the context groups in an OR-relationship.

In general, the software engineer must understand the context, what it is, and how it behaves. Then, during the eCFM modeling, the engineer can reason over how such contexts affect the DSPL. This information is available from the domain analysis.

## 2.5 Optimization methods

Optimization consists of a principle underlying the analysis of complex decision problems. It involves the selection of values for many interrelated variables, by focusing on an objective designed to quantify performance and measure the quality of the decision. Depending on the formulation, such an objective is maximized (or minimized) subject to the constraints that may limit the selection of decision variable values [48]. In general, an objective function consists of a way of assigning value to a possible solution that reflects its quality on a scale. Conversely, a constraint represents a binary assessment reporting whether or not a given requirement contains solutions to a problem in terms of optimization [22]. Optimization, then, should be regarded as a tool for conceptualization and analysis. The problem formulation always involves a trade-off between the conflicting objectives of *(i)* building a mathematical model sufficiently complex to accurately capture the problem description and *(ii)* building a tractable model. Thus, skill in modeling, capturing the essential elements of a problem, and good judgment in the interpretation of results are required to obtain meaningful conclusions [48].

In this first release of our approach, we used ILP as an optimization method, since it is one of the most popular modeling techniques for studies based on simulations. **ILP** is a method for optimization problems that can be applied in a large number of applications such as, in feature selection by helping application engineers in the product configuration activity [24], in an airline wishes to schedule its flight crews, and in an oil company that wants to decide where to drill for oil. This means that, if we can specify the objective as a linear function of certain variables and constraints on resources as equalities or inequalities on those variables, then we have an ILP problem [15]. ILP solves a series of linear equations to satisfy the conditions of the problem while optimizing an objective function. The problem is mathematically designed to find a set of non-negative integer variables, denoted by $x = \{x_1, x_2, ..., x_n\}$, to maximize or minimize a linear objective function of x, denoted by $f(x) = f(x_1, x_2, ..., x_n)$, subject to a set of linear constraints of x, denoted by $c(x) = \{c_1(x_1, x_2, ..., x_n), c_2(x_1, x_2, ..., x_n), ..., c_m(x_1, x_2, ..., x_n)\}$. Any setting of those variables $x$ that satisfies all the constraints is named as a feasible solution to the ILP [15, 48, 79]. The simplex algorithm is the most effective in solving ILP problems. It uses deterministic rules to solve such problems. This algorithm moves from vertex to vertex of the primal feasible region until it reaches an optimal solution and each vertex corresponds to a basic feasible solution [75].

## 3 RELATED WORK

Among the DSPL engineering activities, the variability modeling activity is one of the most important, since it guides the software engineer to handle the diversity of contexts that influence the system's dynamic adaptations [12]. Indeed, different approaches are supporting DSPL variability modeling [2, 38, 68]. From a variability modeling point of view, it is essential to provide support for the configuration selection process and conduct trade-off analysis in DSPL engineering.

Table 1 shows a comparison between the insights found in related work and our approach (ToffA-DSPL). We inserted the studies presented by Hallsteinsen *et. al.* [35], Esfahani *et. al.* [23], Greenwood *et. al.* [30], Guedes *et. al.* [31], Nascimento *et. al.* [55], Sanchez *et. al.* [69], Goldsby *et. al.* [29], Parra *et. al.* [57], Sawyer *et. al.* [71], *et. al.* [58], Ali *et. al.* [3], and Gamez *et. al.* [27, 78] since they consider the trade-off between contextual information and NFRs.

Among them, we ticked which ones use *Utility-based* planning. The table cells (Table 1) that are highlighted in a different color depict such studies, which are presented by Hallsteinsen *et. al.* [35], Esfahani *et. al.* [23], Greenwood *et. al.* [30], Guedes *et. al.* [31], Nascimento *et. al.* [55], and Sanchez *et. al.* [69]. The *Utility-based* planning is considered a suitable strategy when the software engineers need to express the priorities of users over services provided by an application [35]. The priorities, in turn, are represented as weights in the utility function aiming to direct the choice of an optimal solution. Esfahani *et. al.* [23], Greenwood *et. al.* [30], Nascimento *et. al.* [55], and Sanchez *et. al.* [69], for example, propose the usage of this strategy to perform trade-off analysis at runtime. Nevertheless, our approach supports the trade-off analysis at design time as well as the works presented by Hallsteinsen *et. al.* [35] and Guedes *et. al.* [31]. Therefore, it aims to assist software engineers during the initial modeling phase.

Table 1. Related work summary

| Approach | Information | | Strategy | Modeling Characteristics | | |
|---|---|---|---|---|---|---|
| | NFRs | Context | Utility-based planning | Satisfaction Level | Prioritization | Contribution |
| Hallsteinsen *et. al.* [35] | ● | ● | ● | ○ | ● | ○ |
| Esfahani *et. al.* [23] | ● | ● | ● | ● | ○ | ○ |
| Greenwood *et. al.* [30] | ● | ● | ● | ○ | ● | ○ |
| Guedes *et. al.* [31] | ● | ● | ● | ● | ● | ○ |
| Nascimento *et. al.* [55] | ● | ● | ● | ○ | ○ | ○ |
| Sanchez *et. al.* [69] | ● | ● | ● | ○ | ○ | ○ |
| Goldsby *et. al.* [29] | ● | ● | ○ | ● | ● | ○ |
| Parra *et. al.* [57] | ● | ● | ○ | ○ | ○ | ○ |
| Sawyer *et. al.* [71] | ● | ● | ○ | ● | ○ | ○ |
| Pascual *et. al.* [58] | ● | ● | ○ | ● | ○ | ○ |
| Ali *et. al.* [3] | ● | ● | ○ | ○ | ○ | ● |
| Gamez *et. al.* [27] | ● | ● | ○ | ○ | ○ | ○ |
| Welsh *et. al.* [78] | ● | ● | ○ | ● | ● | ○ |
| ToffA-DSPL | ● | ● | ● | ● | ● | ● |

●: Included; ○: Not included

Aiming to identify the approaches that have demonstrated a likely similarity with our approach, we decided to use the following modeling characteristics as a comparison criterion. Such modeling characteristics are employed to support software engineers in decision-making.

- The **satisfaction level** criterion aims to identify whether the authors consider, in their approach, the degree to which each of the variable features satisfies the *soft goals*;
- The **prioritization** criterion verifies whether the approach under evaluation uses the relevance degree of *contexts*, *goals*, and *soft goals*; and
- The **contribution** criterion evaluates whether the authors use, in their approach, the impact of features over *contexts*, *goals*, and *soft goals*.

Hallsteinsen *et. al.* [35] reported conceptual discussions about how to build DSPL projects based on the approach named MADAM. It uses annotations to reason about how well a variant of the DSPL meets its context. For this purpose, the NFRs provided by the DSPL application are compared to those required by the user and those provided by such variants. The match to the user's needs is expressed in a utility function and is used to direct the adaptation. The utility function represents a weighted mean of the differences between the NFRs provided by the DSPL application and the user's preferences over those NFRs. Therefore, the weights represent the priorities of the user and the utility function calculates the benefit of a specific variant of the DSPL. Although this proposal aims to automatically derive changing requirements by monitoring the context and automatically reconfigure the application while it is running, they do not mention the variability model at runtime and how the priorities are measured.

Esfahani *et. al.* [23] provided a framework named Feature-oriented Self-adaptatION (FUSION), which combines feature models with machine learning and in turn improves the accuracy and efficiency of adaptation decisions. Features and NFRs are modeled with the use of goal models. In turn, a *goal* embraces a metric, which is a measurable quantity obtained from the system execution and a utility. The utility function is used to express the user's preferences for achieving a particular metric. In other words, a *goal* defines the user's degree of satisfaction over *soft goal* (*e.g.*, response time) by achieving a specific value of the metric at runtime. The FUSION approach defines several learned functions to estimate the impact of selecting a specific set of features by considering the metrics in a given execution context.

Greenwood *et. al.* [30] presented the DiVA approach, which provides a tool-supported methodology for managing dynamic variability in applications by using the DSPL engineering processes. This approach considers the specific context to which each variation is applicable, as well as, how each variant of the DSPL affects the system and its NFRs. The DiVA approach uses model-driven techniques to model these variability elements and formalize *how* and *when* the system should adapt. In addition, it combines the strengths of both strategies, *ECA rules* and *utility-based* planning aiming to achieve efficiency, scalability, and verification of capabilities. Then, the adaptation rules are expressed as high-level *goals* to achieve and the configuration is optimized concerning these *goals* at runtime. These rules are defined using expressions to describe the *context* that they apply and a set of priorities assigned to the NFRs. In addition, the utility functions are defined to determine how well-suited a configuration is, depending on the *context*.

Guedes *et. al.* [31] proposed an approach called ConG4Das that captures the variability of adaptive systems. It is based on a goal model to represent information such as context, NFRs, the relationship between them, and their priority. Regarding prioritization, such an approach allows a given *context* to be ranked according to the priority of NFRs. It means that the *contexts* affect the required satisfaction level of NFRs.

Nascimento *et. al.* [55] proposed a DSPL infrastructure, called ArCMAPE to support a family of software fault tolerance techniques based on design diversity and instantiates the most suitable one through dynamic variability management. When a requisition is sent to ArCMAPE, the adaptation logic intercepts the running system. In turn, the dynamic adaptation satisfies the rules by maximizing the utility value, which is measured based on pre-defined weights for NFRs. The new configuration is chosen following input values provided by sensors and behavioral changes in the running system. Therefore, the adaptation is triggered by contextual changes or changes in NFRs.

Sanchez *et. al.* [69] proposed an approach for the specification, measurement, and optimization of NFRs based on feature models. It shows how NFRs can be specified utilizing feature attributes by quantitatively evaluating the trade-off among multiple NFRs to arrive at a better system configuration. Their approach requires the mapping of the contexts, events to feature models, the specification of values for NFRs, quality metrics, and weights for the optimization steps. This approach focuses on the quantification of individual attributes and trade-offs among these different metrics, formalizing the problem as the optimization of an objective function that aggregates these metrics and quantifies stakeholders' preferences for individual attributes.

The aforementioned studies provide important information about the configuration selection process of DSPL by considering the trade-off between contexts and NFRs. Among those approaches that assume *Utility-based* planning as a strategy to formalize the knowledge obtained, only Esfahani *et. al.* [23] and Guedes *et. al.* [31] use *goals* to model *how* and *why* the system operates. Both assess the satisfaction level of *soft goals* by varying according to the *context* changes. Conversely, our approach measures the degree to which the variable features satisfy the *soft goals*. In addition, it evaluates the impact of variable features over *contexts*, *goals*, and *soft goals*. Therefore, *contexts* and *soft goals* elements are handled independently.

Additionally, the approach proposed by Guedes *et. al.* [31] does not deal with model integrity constraints according to the relationships between feature and context. In contrast, we propose the eCFM technique to improve context variability expressiveness by specifying real-world constraints related to contextual information.

Some of the main benefits of the eCFM are *(i)* it allows to model constraints among **context features** increasing the *context* variability expressiveness; and *(ii)* the **context group** concept allows to organize the **context features** in different categories defined by the software engineer (*e.g.*, based on purposes or *context* source) favoring the model organization and comprehensibility.

In Figure 2, for instance, one of the functional requirements of the self-adaptive wireless sensor network is represented by the system's feature `Transmit Data` that controls the type of wireless communication. This feature has two variations with an XOR-group: `Bluetooth` and `Wifi`. The feature `Bluetooth` has an exclude dependency relationship with the **context feature** `Alert`. So, it specifies an adaptation rule that is triggered by the **context feature** `Alert`.

We also use the goal model to represent variable features and the set of required NFRs for satisfying stakeholders' intentions. Nevertheless, both eCFM and the goal model are integrated to support the software engineer in the decision-making process. Since we provide an approach that takes into account the combination of several modeling characteristics such as *satisfaction level*, *prioritization*, and *contribution*, it is possible to promote a more accurate representation of how DSPL applications should operate in real-world environments.

## 4 THE APPROACH

Essentially, we propose the ToffA-DSPL approach to manage both dimensions *structural variability* and *context variability*. It is addressed to the modeling of the system's features and contexts by meeting specific quality requirements. In this sense, it is composed of eCFM and goal model as key references to assist communication with stakeholders. In addition, the approach is used to support the configuration selection process in DSPL engineering by considering that contexts can influence the way of satisfying the NFRs of each model variant and vice versa [2].

We argue that the specification of DSPL can be done by using the approach proposed in this section since it facilitates software engineers to achieve consensus with stakeholders and understand their preferences and needs. In this sense, the ToffA-DSPL approach aims to support trade-off analysis and can be reduced to an optimization model composed by a *utility function*. This strategy can express users' priorities *over* services provided by an application [35]. Additionally, it uses a solver based on the ILP technique [33] to run the configuration process. As a result, it is possible to identify optimal configurations that better meet stakeholders' preferences, the variability of system features, contexts, NFRs, and constraints.

Figure 3 shows the steps proposed for supporting the configuration selection process in DSPL engineering. We explain it through a running example in the self-adaptive wireless sensor network domain named *GridStix* [29, 42, 71], depicted in Figure 2. It establishes an automated mechanism to warn about floods on rivers. We illustrate the approach with this running example within boxes along this section.

### 4.1 Domain analysis

**STEP 1–Domain analysis** is first performed by the software engineer to identify the following main elements for modeling DSPL, as follows:

- **Functional requirements** represented as system features;
- **Variability information** specified as feature relationships of the types mandatory, optional, OR-group or XOR-group;
- **Context information** relevant to the DSPL;
- **Adaptations rules**, which are specified as dependencies among features and contexts;
- **Stakeholder's preferences**, which are defined based on the system's features;
- **Variability space**, which consists of the set of variable features that influence the decision-making process; and

Fig. 3. The DSPL Trade-off Analysis (ToffA-DSPL) approach

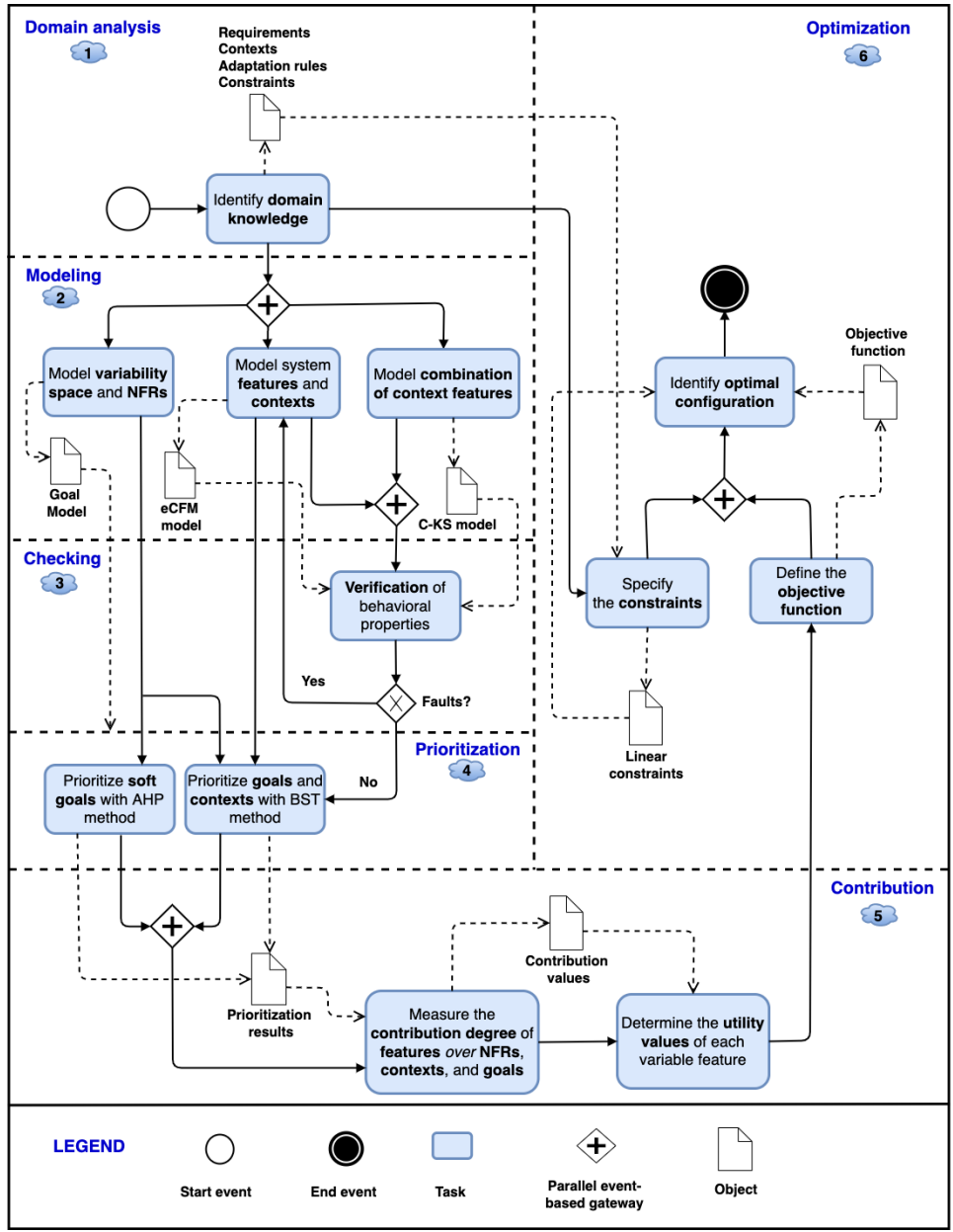

- **A set of required NFRs** that should be defined by the software engineer with the support, for instance, of a catalog such as the one proposed by Uchoa *et al.* [76].

## 4.2 Modeling

**STEP 2–Modeling** comes next. The software engineer should model the DSPL using eCFM and goal model techniques. The eCFM should be specified before the goal model since *goals* and *hard goals* are defined based on the system's features. For building such a model, the software engineer should use the knowledge acquired at **STEP 1**, as follows:

- **Functional requirements** and the **variability information** leads to the DSPL features and their relationships;
- **Contexts information** are used to define **Context Root**, **Context Groups**, and **Context Features**;
- **Adaptation rules** are represented in accordance with the require and exclude relationships. In the first relationship, the feature strongly satisfies the context, whereas, in the second, the feature is strongly denied by the context; and
- The dependencies between **context features** with their respective **context groups** should be represented in accordance with XOR-group, OR-group or optional relationships. For the XOR-group relationship, two or more **context features** cannot occur at the same time, *i.e.*, only one **context feature** is mandatory to satisfy its **context group**. In contrast, for the OR-group and optional relationships, all **context features** may satisfy their **context group**.

> The running example encompasses four control systems, as follows:
> - **Transmit data** (mandatory feature) is composed by two alternative features such as **Bluetooth** and **Wifi** for internode data transmission;
> - **Organize network** (mandatory feature) is composed by two alternative features such as **FHTopology** and **SPTopology** for routing data between nodes;
> - **Calculate flow rate** (mandatory feature) is composed of two alternative features such as **Single Node Processing** and **Distributed Processing** to facilitate the operation of nodes for extended periods; and
> - **Measure depth** (mandatory feature) of water.
>
> Both **State of River** and **Health of Battery** were identified as **context groups**, which can influence possible adaptations. Each one is composed of three and two **context features**, respectively. For instance, whether contexts **Low** and **Emergency** are detected in the environment execution, the feature **Bluetooth** may be activated (require relationship). In contrast, the feature **Wifi** may be deactivated (exclude relationship). The same reasoning can be applied to all adaptation rules.
>
> In the running example, we consider that all **context features** have XOR-group relationship with their respective **context group**. For instance, contexts $c_7$ (**High**) and $c_8$ (**Low**) cannot occur at the same time in a real-world environment.

The goal model (see Figure 4) encompasses high-level goals, hard goals, soft goals, and mapping links among them. For building such a model, the software engineer should use the variability space, the set of required NFRs, and stakeholder's requirements identified at **STEP 1**. The mapping link between goals and hard goals represents the AND/OR relationships, as follows:

- The AND relationship means that all hard goals are mandatory to satisfy the goal;
- The OR relationship means that each hard goal can satisfy its goal; and

Fig. 4. Grid Stix modeled with Goal Model. It represents goals, hard goals, soft goals, and their respective relationships.

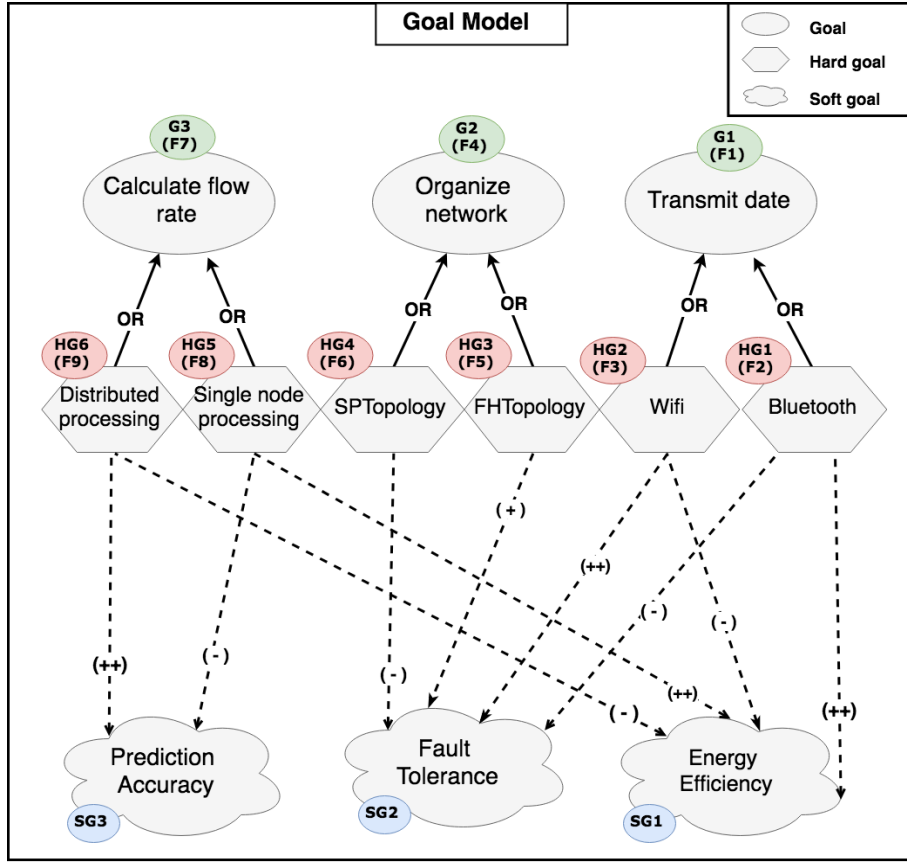

- The mapping link between hard goals and soft goals defines the satisfaction level: **satisfied** (++) = 1, **weakly satisfied** (+) = 0.5, **undecided** (?) = 0, **weakly denied** (-) = -0.5, and **denied** (- -) = -1.

> Figure 4 shows the goal model of the running example, which is composed of three soft goals: **Energy Efficiency** ($sg_1$), **Fault Tolerance** ($sg_2$), and **Prediction Accuracy** ($sg_3$). In addition, we defined the following goals:
> - **Transmit date** ($g_1$), which includes the hard goals **Bluetooth** and **Wifi** ($hg_1$ and $hg_2$);
> - **Organize network** ($g_2$), which includes the hard goals **FHTopology** and **SPTopology** ($hg_3$ and $hg_4$); and
> - **Calculate flow rate** ($g_3$), which includes the hard goals **Single node processing** and **Distributed processing** ($hg_5$ and $hg_6$).

The third model, named *Context Kripke Structure* (C-KS), concerns the possible combination of context features (see Definition 4.1). Hence, it specifies the relationship among context features (*e.g.*, indicating contexts that cannot be true at the same time).

DEFINITION 4.1 (COMBINATION OF CONTEXT FEATURES). *Given a set of context groups, a CCF is a combination of context features that can be detected in the environment at runtime.*

The C-KS model is build from the context information gathered in **STEP 1**. Indeed, it is a state-based formalism that captures the context variation over time by modeling the context variations as a transition graph in which nodes are context states and edges are transitions that represent changes in context [63]. In this work, the states of C-KS are represented by CCFs identified from the feature model.

> By observing the eCFM presented in Figure 2, we identified six possible CCFs due to XOR-group relationship between **context feature** and its respective **context group**. As depicted in Table 2, for instance, $ccf_1$ includes the context features **Emergency** ($c_3$) and **High** ($c_7$). It means that such context features can be detected at the same time during application execution. In this running example, the following constraints among the context features were considered:
> - From the context feature **Normal** ($c_4$), next river state is **Alert** ($c_5$);
> - From the context feature **Alert** ($c_5$), next river state can be **Normal** ($c_4$) or **Emergency** ($c_3$); and
> - From the context feature **Emergency** ($c_3$), next river state can be only **Alert** ($c_5$).
>
> Figure 5 presents the C-KS model encompassing these constraints and specifying the possible transitions among the CCFs. In the running example, we have the following transitions: (i) from $ccf_1$ the next possible CCFs are $ccf_3$ or $ccf_4$; (ii) from $ccf_2$ the next CCFs are $ccf_3$ or $ccf_5$; (iii) from $ccf_3$ the next CCFs are $ccf_2$, $ccf_1$ or $ccf_6$; (iv) from $ccf_4$ the next CCFs are $ccf_1$ or $ccf_6$; (v) from $ccf_5$ the next CCFs are $ccf_2$ or $ccf_6$; and (vi) from $ccf_6$ the next possible CCFs are $ccf_3$, $ccf_4$ or $ccf_5$.

Table 2. CCFs of the running example Grid-Stix

| CCF | Contexts |
|---|---|
| $ccf_1$ | **Emergency** ($c_3$), **High** ($c_7$) |
| $ccf_2$ | **Normal** ($c_4$), **High** ($c_7$) |
| $ccf_3$ | **Alert** ($c_5$), **High** ($c_7$) |
| $ccf_4$ | **Emergency** ($c_3$), **Low** ($c_8$) |
| $ccf_5$ | **Normal** ($c_4$), **Low** ($c_8$) |
| $ccf_6$ | **Alert** ($c_5$), **Low** ($c_8$) |

### 4.3 Checking

The goal of the **STEP 3 – Checking** is to check for design faults in the DSPL specification. Our approach follows the DSPL model checking technique depicted in [70]. This verification aims to avoid unexpected behavior of the application at runtime.

First, from both e-CFM and C-KS models the software engineer should design the DAS behavior in Promela [39], which is a verification modeling language. From the e-CFM, the software engineer identifies the features, context, adaptation rules, and constraints among features. The C-KS, in turn, provides information about the contextual variation, allowing the model checking to reduce the number of possible context states to be explored and verify real scenarios by avoiding the identification of false-positive faults. Hence, the software engineer can check behavioral proprieties over the DAS by using the Spin model checker tool[1].

Fig. 5. C-KS of the running example

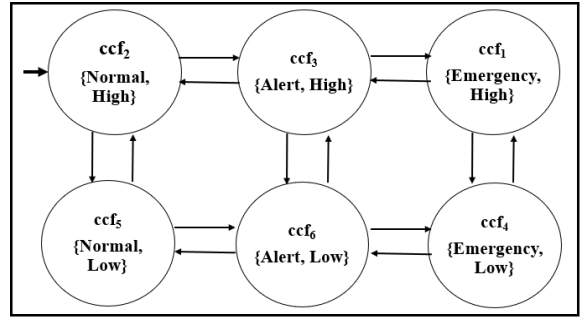

Last, we suggest checking the five properties presented in [70]. If faults are identified in the eCFM, the software engineer should fix the model and check the properties again until no more design faults are detected.

> In the running example, the C-KS was specified as a process named ContextManage, which is in charge to enable the context changes. Regarding the contexts (*e.g*, **Emergency** and **Normal**) and features (*e.g*, **Wifi** and **FhTopology**), they were defined as variables with the value true (currently active) or false (currently not active).
>
> The adaptation rules were defined in Promela according to Figure 2. In this case, for example, one of the adaptation rules states that if the river state is **Emergency**, the features **Single Node** and **Bluetooth** should be activated. Last, for the running example, we checked the properties defined by Santo et al. (2016).
>
> After running the check of each property, most of them returned no error. However, we identified an interleaving fault due to the $ccf_1$ (**Emergency**, **High**). Such context features trigger rules that interact with each other, for instance, context feature **Emergency** deactivates feature **Distributed Processing** ($f_9$) while context feature **High** activates this feature.

## 4.4 Prioritization

Since all modeling elements are already captured and modeled, then goals, soft goals, and contexts should be prioritized according to the stakeholder's preferences. The prioritization results can be used as parameters to measure the contribution of the system's features (hard goals) connected to the contexts, goals, and soft goals.

The **STEP 4 – Prioritization** consists of the relevance degree of *goals*, *contexts*, and *soft goals*. Two ranking and prioritization methods are used: Analytical Hierarchy Process (AHP) [66] and Binary Search Tree (BST) [45]. We evidenced that these ranking and prioritization methods are used to better meet stakeholders' preferences. With the AHP method, for instance, it is possible to check the consistency of the results based on a ratio scale. Indeed, the AHP method brings a scalability problem for larger projects. However, it can be reduced with the use of tools to support the configuration process described in this work [45]. The combination of both methods, BST and AHP help us to communicate with stakeholders and identify the potential preferences in terms of system quality.

---
[1] http://spinroot.com/spin/Src/index.html

- **Prioritization of goals** – The BST method was used to rank the selected goals, which is an efficient method for prioritizing large-scale items [45]. In this method, each goal is represented by a node of the tree and has sub-nodes, which can be *subgoals* or *hardgoals*. The tree is organized according to priorities of *goals* and *hardgoals*. Then, such elements must be ordered from right to left in order to specify the sub-tree. It means that the right side of the sub-tree contains requirements with higher priority than the left side of the sub-tree. The goal prioritization can be conducted according to the process described by [60], as follows:

We identify at the variability space, *goals* and *hard goals* related to functional requirements. Next, we selected a *goal* with a higher priority and put it as a root node. After that, another goal $g_b$ can be selected and compared with the root *goal* in terms of its importance. If $g_b$ has a lower priority than the root node $g_a$, it needs to be compared with the left sub-node and so forth. This process should be repeated until all *goals* have been compared and inserted into the BST. This same process should be made with *hard goals*.

> First, three high-level goals related to functional requirements ($g_1$ – **Transmit date**, $g_2$ – **Organize network**, and $g_3$ – **Calculate flow rate**) and six hard goals ($hg_1$ – **Bluetooth**, $hg_2$ – **Wifi**, $hg_3$ – **FHTopology**, $hg_4$ – **SPTopology**, $hg_5$ – **Single node processing**, and $hg_6$ – **Distributed processing**) have been identified. Next, each goal, starting with $g_1$, was compared with all others (Figure 4) and the following order or priority has been established $g_2$, $g_1$, and $g_3$.

Then, a normalized rank value is assigned to each goal according to the formula:

$$rankValue = \frac{1}{1 + [rank]} \qquad (1)$$

The rank is a series of crescent natural numbers starting from 1. Thus, the *rankValue* may be transformed into a scale ranging from 0 (exclusive) to 1 (inclusive), where the value close to zero means the lowest priority and the value close to one means the highest priority.

> Table 3 shows the rank value for each *goal* and the related normalized *rankValue* of the running example.

- **Prioritization of contexts** – It used the same steps aforementioned described for context prioritization.

> Following the running example, two high-level contexts needed to be prioritized ($c_2$ – **State of river** and $c_6$ – **Health of Battery**). Starting with $c_2$ we compared it with $c_6$ in terms of its importance and defined the following order: $c_2$ and $c_6$. After that, we calculated the rank values and normalized them as shown in Table 3.

Table 3. Prioritization using the BST method

|  | Goal | | | Context | |
|---|---|---|---|---|---|
|  | $g_2$ | $g_1$ | $g_3$ | $c_2$ | $c_6$ |
| *rank* | 1 | 2 | 3 | 1 | 2 |
| *rankValue* | 0.5 | 0.33 | 0.25 | 0.5 | 0.33 |

- **Prioritization of soft goals** – We can determine the priority of soft goals according to the stakeholder's preferences using the AHP method. It is also based on a pairwise comparison process to generate a ranked list of soft goals. In addition, it evaluates and checks the consistency of judgments.

  Assuming $SG$ as a set of soft goals, a comparison matrix $A[n, n]$ must be created to show the relative importance of each pair of soft goals. The soft goals are compared against each other according to the scale of importance: 1 (**equal**), 3 (**moderate**), 5 (**strong**), 7 (**very strong**), and 9 (**extreme**), respectively [66]. If the soft goal in column $j$ is preferred to the soft goal in row $i$, then the inverse of the rating is given ($a_{j,i} = \frac{1}{a_{i,j}}$), *i.e.*, we put the actual judgment value on the right side of the matrix diagonal row and the reciprocal value in the left side of the diagonal. In the running example, Table 4 shows the complete comparison matrix $A[3, 3]$. It shows the relative importance of each pair of soft goals.

  The next step is to normalize the comparison matrix and calculate the importance value ($iValue$) for each soft goal. Totaling the numbers in each column does this step. Each entry in column $j$ is then divided by the column $sum$ to yield its normalized score. As a result, the $sum$ of each column is 1. The $iValue$ is calculated as follows:

  $$iValue = \frac{\sum_{i=1}^{n} A[i, j]}{n} \quad (2)$$

  Table 5 shows the normalized matrix of the running example and the $iValue$ for each soft goal. The $iValue$ is used to measure the contribution of features over NFRs.

Table 4. Relative importance matrix $A[n, n]$

|     | $sg_1$ | $sg_2$ | $sg_3$ |
|-----|--------|--------|--------|
| $sg_1$ | **1** | 3 | 3 |
| $sg_2$ | 0.33 | **1** | 1 |
| $sg_3$ | 0.33 | 1 | **1** |
| $sum$ | 1.66 | **5** | **5** |

Table 5. The normalized matrix

|     | $sg_1$ | $sg_2$ | $sg_3$ | $iValue$ |
|-----|--------|--------|--------|----------|
| $sg_1$ | 0.6 | 0.6 | 0.6 | **0.6** |
| $sg_2$ | 0.2 | 0.2 | 0.2 | **0.2** |
| $sg_3$ | 0.2 | 0.2 | 0.2 | **0.2** |
| $sum$ | 1 | 1 | 1 | **1** |

The software engineer can decide between different and equal priorities. For instance, whether the soft goals $sg_1$, $sg_2$, and $sg_3$ have the same priority, their rank values are equal to 1. The same reasoning can be applied to all goals and contexts.

### 4.5 Contribution

**STEP 5 – Contribution** consists of the impact of features over *goals*, *contexts*, and *soft goals*. It is performed after prioritizing all the model elements.

- **Features over goals** – We use a diagrammatic reasoning approach introduced by Ali *et al.* [2] to calculate the contribution degree of features over *goals* (equation 3). It shows that top-level goals are iteratively decomposed into sub-goals by AND–relationship and OR–relationship. Such goals are satisfied by means of executable tasks (*i.e.*, hard goals), as follows:

$$Cont(f_i, g_i) = rankValue(g_i) \times satValue(hg_i) \tag{3}$$

Where $rankValue(g_i)$ shows the priority value of a goal and $satValue(hg_i)$ shows to what extent each hard goal can satisfy its goal. We calculate $satValue(hg_i)$ based on the AND/OR relationships. In the AND relationship, $satValue(hg_i)$ is divided by the number of hard goals $m$. Otherwise, in an OR relationship, the satisfaction value is 1, since each hard goal can satisfy its goal.

> In the running example, the hard goals $hg_1$ (feature $f_2$) and $hg_2$ (feature $f_3$) have an OR relationship with their parent goal $g_1$ resulting in $satValue(hg_1) = 1$ and $satValue(hg_2) = 1$. Thus, the impact of features $f_2$ and $f_3$ over goal $g_1$ can be computed, as follows:
> (iv)
> $$Cont(f_2, g_1) = satValue(hg_1) \times rankValue(g_1)$$
> $$= 1 \times 0.33 = 0.33$$
> (v)
> $$Cont(f_3, g_1) = satValue(hg_2) \times rankValue(g_1)$$
> $$= 1 \times 0.33 = 0.33$$

- **Features over contexts** – The same reasoning used when handling features over goals can be applied to the decomposition of *contexts*. Therefore, we also use a diagrammatic reasoning approach introduced by Ali *et al.* [2] to calculate the contribution degree of features over *contexts* (equation 4). It shows that **context groups** are decomposed into **context features** by AND–relationship and OR–relationship. Such **context groups** are satisfied by means of **context features**, as follows:

$$Cont(f_i, c_{fi}) = \sum_{\forall c \in C | f \rightarrow c} rankValue(c_{gi}) \times satValue(c_{fi}) \times impDegree(c_{fi}) \tag{4}$$

The function $rankValue$ calculates the priority value of a **context group** ($c_{gi}$). The function $satValue(c_{fi})$ shows to what extent each **context feature** ($c_{fi}$) can satisfy its **context group** by considering the AND/OR relationships. In the AND relationship $satValue(c_{fi})$ is divided by the number of **context features m**. Otherwise, in an OR relationship the satisfaction value is 1 since each **context feature** can satisfy its **context group**. The function $impDegree$ displays to what extent each feature can satisfy a **context feature** in the eCFM according to the require and exclude relationships between **context feature** and system's feature. The first is represented by an impact degree with value **1**, meaning that the feature strongly satisfies the **context feature**. The second is represented by an impact degree with value **-1**, meaning that the feature is strongly denied by the **context feature**.

> We identify the AND/OR relationships of each **context feature** ($c_{fi}$) with its respective **context group** ($c_{gi}$). Following the running example, we consider that all **context features** have OR–relationship with their respective **context group**, resulting in $satValue(c_{fi}) = 1$. Thus, feature $f_2$ for instance, has contribution degree over *contexts* $c_3$ and $c_8$, as follows:
> (i)
> $$Cont(f_2, c_3) =$$
> $$rankValue(c_2) \times satValue(c_3) \times impDegree(c_3)$$
> $$= 0.5 \times 1 \times 1 = 0.5$$

(ii)
$$Cont(f_2, c_8) = \\ rankValue(c_6) \times satValue(c_8) \times impDegree(c_8) \\ = 0.33 \times 1 \times 1 = 0.33$$

(iii)
$$Cont(f_2, c_3) + Cont(f_2, c_8) = 0.5 + 0.33 = 0.83$$

- **Features over soft goals** – Contribution of features over *soft goals* (equation 5) can be identified by the mapping links in Figure 4 and calculated as follows:

$$Cont(f_i, sg_i) = \sum_{\forall sg \in SG | f \to sg} iValue(sg_i) \times impDegree(sg_i) \tag{5}$$

Where $iValue$ is the importance value of soft goals and $impDegree(sg_i)$ shows to what extent each feature can satisfy a soft goal in the goal model, based on the conversion schema for satisfaction level: $(--) = -1$, $(-) = -0.5$, $(?) = 0$, $(+) = 0.5$, $(++) = 1$.

Following the running example, feature $f_2$ ($hg_1$) is related to soft goals $sg_1$ and $sg_2$, as follows:

(vi)
$$Cont(f_2, sg_1) = iValue(sg_1) \times impDegree(sg_1) \\ = 0.6 \times 1 = 0.6$$

(vii)
$$Cont(f_2, sg_2) = iValue(sg_2) \times impDegree(sg_2) \\ = 0.2 \times (-0.5) = -0.1$$

(viii)
$$Cont(f_2, sg_1) + Cont(f_2, sg_2) = 0.6 + (-0.1) = 0.5$$

The **utility values** $\mathbb{C}_{fi}$ of each variable feature $f_i$ can be determined by evaluating the contribution degree of its associated *hard goals* over *goals* and *soft goals*, besides of the contribution degree of its associated *hard goals* over *contexts* (equation 6), as follows:

$$\mathbb{C}(f_i) = \sum_{u=1}^{|G|} Cont(f_i, g_u) + \sum_{t=1}^{|SG|} Cont(f_i, sg_t) + \sum_{v=1}^{|C|} Cont(f_i, c_v) \tag{6}$$

Table 6 presents the utility values for the running example that are measured, as follows:

*(i)* The sum of contribution values for feature $f_2$ over contexts $c_3$ and $c_8$ resulted in a *utility value* equal to 0.38 (equation 3). This process should be repeated for all features and their respective **context features** to complete the first column of the table;

*(ii)* Since the features $f_2$ and $f_3$ only contribute *over* goal $g_1$, the resulting *utility values* are equal to 0.5 and 0.5 (equations 4 and 5). This process should be repeated for all features and their respective goals to complete the second column of the table; and

*(iii)* The sum of contribution values for feature $f_2$ over soft goals $sg_1$ and $sg_2$ resulted in a *utility value* equal to 0.5 (equation 8). This process should be repeated for all features and their respective soft goals to complete the third column of the table.

Table 6. Utility value for features by considering Goal, Soft goal, and Context (Configuration $F_1$)

| System feature | $Cont(fi, Cs)$ | $Cont(fi, Gs)$ | $Cont(fi, SGs)$ | $\mathbb{C}(fi)$ |
|---|---|---|---|---|
| $f_2$ | 0.83 | 0.33 | 0.5 | 1.66 |
| $f_3$ | -0.33 | 0.33 | -0.1 | -0.1 |
| $f_5$ | 0 | 0.5 | 0.1 | 0.6 |
| $f_6$ | 0 | 0.5 | -0.1 | 0.4 |
| $f_8$ | 0.83 | 0.25 | 0.5 | 1.58 |
| $f_9$ | -0.83 | 0.25 | -0.1 | -0.68 |

The utility values are the coefficients of the decision variables. The utility value $\mathbb{C}_{fi}$ will be used in the optimization model.

### 4.6 Optimization

After executing the previous steps, the software engineer should perform the **STEP 6 – Optimization**. For that, we used the utility function as a strategy to deal with trade-off analysis. Based on the utility values, we defined an optimization model that suggests valid and optimal configurations by considering the integrity constraints and variability of the feature model. In addition, it ensures that solutions satisfy *contexts*, *goals*, *soft goals*. The optimization model is characterized, as follows:

(1) A set of **decision variables** $X_{fi}$ whose value is 1 if the feature $x_{fi}$ is active, 0 otherwise;

(2) An **objective function** that measures the decision variables summation by satisfying the *goals*, *soft goals*, and *contexts* represented in eCFM and the goal model (equation 7). The result of the problem formulation considers the utility value $\mathbb{C}_{fi}$ of the system features (*hard goals*), as follows:

$$\max \sum_{fi}^{n} \mathbb{C}_{fi} \cdot X_{fi}, \forall f_i \in F \tag{7}$$

> In the running example, features that are not variable and are not represented as hard goals in the goal model receive value 0 as a coefficient to eliminate their impact on the maximization of the objective function, as follows:
>
> $$\max\ 0.0X_{f0} + 0.0X_{f1} + 1.66X_{f2} - 0.1X_{f3} + 0.0X_{f4} + 0.6X_{f5} + 0.4X_{f6} + 0.0X_{f7} + 1.58X_{f8}$$
> $$- 0.68X_{f9} + 0.0X_{f10}$$

(3) A set of **linear constraints** that are subject to the variability and integrity constraints of eCFM and the goal model (equations 8, 9, 10, 11, 12, and 13). They are based on the relationship model described by Kang *et al.* [44], as follows:

Let $F$ be a set of features, $|F| = n$ and $F^m$ be a feature model that represents a hierarchical relationship between features. Thus, the ordered pair of features $(f_p, f_c) \in F^m$ if $f_c$ is a child of $f_p$.

Let $M \subseteq F^m$ be the set of pair of features with mandatory relation. If $(f_p, f_c) \in M$, then both features $f_p$ and $f_c$ must be activated or deactivated at the same time:

$$x_{fc} = x_{fp}, (f_p, f_c) \in M \tag{8}$$

The optional relationship denotes:
$$x_{fc} \leq x_{fp}, (f_p, f_c) \in F^m \tag{9}$$

The set $O \subseteq F^m$ denotes the OR relation:
$$\sum x_{fc} \geq x_{fp}, (f_p, f_c) \in O, \forall f_c \in F \tag{10}$$

The set $A \subseteq F^m$ denotes the alternative relation:
$$\sum x_{fc} = x_{fp}, (f_p, f_c) \in A, \forall f_c \in F \tag{11}$$

The set $R \subseteq F \times F$ denotes the require relation between features, thus $(f_r, f_k) \in R$ means that if $f_r$ is activated then $f_k$ must be activated:
$$x_{fr} \leq x_{fk}, (f_r, f_k) \in R \tag{12}$$

The set $E \subseteq F \times F$ denotes the exclude relation between features, thus $(f_e, f_k) \in R$ means that $f_e$ and $f_k$ must not be activated at the same time:
$$x_e + x_k \leq 1, (f_e, f_k) \in E \tag{13}$$

---

In the running example, we defined the constraints, as follows:
- Mandatory relations between features (Constraint 1): $X_{f1} = X_{f0}, X_{f4} = X_{f0}, X_{f7} = X_{f0}, X_{f10} = X_{f0}$;
- Considering the features with an alternative relation, if a parent feature is activated only one child will be activated (Constraint 3): $X_{f2} = X_{f1}, X_{f3} = X_{f1}, X_{f5} = X_{f4}, X_{f6} = X_{f4}, X_{f8} = X_{f7}, X_{f9} = X_{f7}$;
- Feature $f_k$ will be deactivated if it has an exclude relationship with $f_e$ (Constraint 6): $X_{f2} + X_{f9} \leq 1$.

---

The approach is partially automated to perform the measurements of prioritization, contribution, and satisfaction level. In addition, we used the TSP library [37] to numerically represent the features and constraints that were defined in the model. Such a library of sample instances is used by a solver named Gurobi [33], which is based on the ILP technique to run the configuration process and find optimal configurations that meet all constraints.

---

For the current example, the time spent in the execution was 0.20 seconds. In addition, the output of the ILP solver suggested that the set of features $F_1 = \{f_1, f_2, \neg f_3, f_4, f_5, \neg f_6, f_7, f_8, \neg f_9, f_{10}\}$ satisfies the soft goals ($sg_1 - sg_3$), goals ($g_1 - g_3$), and the contexts ($c_1 - c_8$) to the aforementioned prioritization. It indicated that features $f_3$, $f_6$, and $f_9$ negatively influence the soft goals of other features, such as *energy save* and *fault tolerance*.

For example, the contribution values in feature $f_9$ (*e.g.*, −0.83 and −0.1) indicate that they negatively influence the contexts and soft goals in DSPL. In addition, it has an exclude relationship with feature $f_2$, which was kept in the optimal configuration. Therefore, the DSPL developer must deal with the integrity constraints in terms of implementation by considering the adaptation rules represented in eCFM.

---

## 5 REASONING ABOUT ADAPTABILITY

In this section, we describe the usefulness of the ToffA-DSPL approach from four different points of view: Firstly, how to identify design faults in the feature model 5.1 Secondly, how to conduct trade-off analysis by considering all the elements that comprise eCFM and goal model (Section 5.2). Thirdly, which optimal configuration meets a specific CCF (Section 5.3). Finally, the approach supports the definition of adaptation models for DSPLs from optimal configurations found in the CCF change analysis (Section 5.4). Such a description of how using the ToffA-DSPL approach is based on gathered data from different simulations.

## 5.1 Design Faults identification

As presented in Section 4.3, the third step of our approach is to perform a model checking to identify possible design faults in the feature model of a DSPL. Identifying these faults can help the software engineer fix problems in the specification to avoid unexpected behavior. For some faults, the fixing may be the adding or excluding of some adaptation rule. Also, it may be necessary to review the captured contexts and their impact on the system's features. At last, when conflicts are detected, they can be solved by the ToffA-DSPL since it provides optimal configurations considering the prioritization of *contexts*, *goals*, and *soft goals*.

After checking behavioral properties (reported in a previous paper [70]), we identified a fault related to interleaving among contexts. In this case, features activated by a context are deactivated by another context of the same CCF, as follows: contexts **Emergency** ($c_3$) and **High** ($c_7$) have an interleaving fault because they trigger actions over the same features, *e.g.*, while context $c_3$ requires the features **Distributed Processing** ($f_9$) and **Wifi** ($f_3$) deactivated, context $c_7$ requires these features activated. It means that feature $f_3$ can be activated (**Wifi** = 1) when context $c_3$ is true.

One solution for this kind of fault is the prioritization of the modeling elements (contexts, goals, and soft goals) and the definition of the order of execution in cases of rules interleaving. For this purpose, a software engineer may perform a trade-off analysis to support this prioritization. The next section discusses this analysis and how it can be done.

## 5.2 Trade-off analysis

Stakeholders' preferences change over time and are hard to elicit. Thus, we propose the trade-off analysis aiming to find valid and optimal configurations that can meet such preferences. Trade-off analysis consists of simulating changes in the prioritization of *goals*, *soft goals*, and *contexts*. For each change, software engineers must only consider relationships between the system's features and the possible CCFs.

Table 7 depicts the optimal configurations found in the trade-off analysis, whereas Table 8 presents for which CCF a specific optimal configuration was suggested by the ILP solver. In our running example, we simulated scenarios that correspond to changes in prioritization for soft goals, goals, and contexts. In each scenario, we made only one prioritization change but always considered the same satisfaction levels as defined in Figure 4. In total, we got six scenarios, of which two were related to prioritization changes in soft goals, two had changes in goals, and others two in context.

The table cells (Table 7) that are highlighted in a different color depict that the modeling element has its prioritization changed. Observing the optimal configurations from prioritization $P_3$ and $P_4$, they indicate that the change of prioritization for *goals* did not affect the results suggested by the ILP solver. We can also notice the same by verifying the Table 8 in $P_3$ and $P_4$. For both prioritizations, the ILP solver suggested $F_1$, $F_2$, $F_3$, and $F_4$ as optimal configurations.

Observing the optimal configurations generated by the simulated scenarios, they indicate that the change of prioritization for *goals* did not affect the results suggested by the ILP solver. Hence, in both scenarios where the goals' prioritization was changed (i.e., g3 > g2 > g1 and g1 > g3 > g2), the resulting optimal configuration suggested by the ILP solver was the same.

In contrast, when we changed the prioritization for soft goals, the ILP solver suggested different configurations. For instance, given the prioritization of soft goals as $sg_2 > sg_3 > sg_1$, and the CCF as $ccf_1$, the ILP solver suggested suggesting as optimal configuration the $F_3$. However, with the prioritization $sg_1 > sg_2 > sg_3$, for the same CCF, the suggested configuration is $F_1$. The main difference between this configuration is due to the status of features $f_2$ and $f_3$. In $F_1$, $f_2$ is activated and $f_3$ is deactivated, while in $F_3$ we have the opposite. So, it means that soft goals can potentially affect the activation and deactivation of the system's features according to their contribution values.

Table 7. An example of trade-off analysis. The columns highlighted in a different color depict that the modeling element has its prioritization changed.

| | | | |
|---|---|---|---|
| | **Change of Prioritization for Soft goals** | | |
| $P_1$ | $sg_2 > sg_3 > sg_1$ | | $F_1 = f_0, f_1, f_2, \neg f_3, f_4, f_5, \neg f_6, f_7, f_8, \neg f_9, f_{10}$ |
| | $c_2 > c_6$ | | $F_3 = f_0, f_1, \neg f_2, f_3, f_4, f_5, \neg f_6, f_7, f_8, \neg f_9, f_{10}$ |
| | $g_2 > g_3 > g_1$ | | $F_4 = f_0, f_1, \neg f_2, f_3, f_4, f_5, \neg f_6, f_7, \neg f_8, f_9, f_{10}$ |
| $P_2$ | $sg_1 > sg_2 > sg_3$ | | $F_1 = f_0, f_1, f_2, \neg f_3, f_4, f_5, \neg f_6, f_7, f_8, \neg f_9, f_{10}$ |
| | $c_2 > c_6$ | | $F_2 = f_0, f_1, f_2, \neg f_3, f_4, f_5, \neg f_6, f_7, \neg f_8, f_9, f_{10}$ |
| | $g_2 > g_3 > g_1$ | | $F_3 = f_0, f_1, \neg f_2, f_3, f_4, f_5, \neg f_6, f_7, f_8, \neg f_9, f_{10}$ |
| | **Change of Prioritization for Goals** | | |
| $P_3$ | $sg_3 > sg_2 > sg_1$ | | $F_1 = f_0, f_1, f_2, \neg f_3, f_4, f_5, \neg f_6, f_7, f_8, \neg f_9, f_{10}$ |
| | $c_6 > c_2$ | | $F_2 = f_0, f_1, f_2, \neg f_3, f_4, f_5, \neg f_6, f_7, \neg f_8, f_9, f_{10}$ |
| | $g_3 > g_2 > g_1$ | | $F_3 = f_0, f_1, \neg f_2, f_3, f_4, f_5, \neg f_6, f_7, f_8, \neg f_9, f_{10}$ |
| | | | $F_4 = f_0, f_1, \neg f_2, f_3, f_4, f_5, \neg f_6, f_7, \neg f_8, f_9, f_{10}$ |
| $P_4$ | $sg_3 > sg_2 > sg_1$ | | $F_1 = f_0, f_1, f_2, \neg f_3, f_4, f_5, \neg f_6, f_7, f_8, \neg f_9, f_{10}$ |
| | $c_6 > c_2$ | | $F_2 = f_0, f_1, f_2, \neg f_3, f_4, f_5, \neg f_6, f_7, \neg f_8, f_9, f_{10}$ |
| | $g_1 > g_3 > g_2$ | | $F_3 = f_0, f_1, \neg f_2, f_3, f_4, f_5, \neg f_6, f_7, f_8, \neg f_9, f_{10}$ |
| | | | $F_4 = f_0, f_1, \neg f_2, f_3, f_4, f_5, \neg f_6, f_7, \neg f_8, f_9, f_{10}$ |
| | **Change of Prioritization for Contexts** | | |
| $P_5$ | $sg_2 > sg_3 > sg_1$ | | $F_3 = f_0, f_1, \neg f_2, f_3, f_4, f_5, \neg f_6, f_7, f_8, \neg f_9, f_{10}$ |
| | $c_2 > c_6$ | | $F_4 = f_0, f_1, \neg f_2, f_3, f_4, f_5, \neg f_6, f_7, \neg f_8, f_9, f_{10}$ |
| | $g_3 > g_1 > g_2$ | | |
| $P_6$ | $sg_2 > sg_3 > sg_1$ | | $F_1 = f_0, f_1, f_2, \neg f_3, f_4, f_5, \neg f_6, f_7, f_8, \neg f_9, f_{10}$ |
| | $c_6 > c_2$ | | $F_3 = f_0, f_1, \neg f_2, f_3, f_4, f_5, \neg f_6, f_7, f_8, \neg f_9, f_{10}$ |
| | $g_3 > g_1 > g_2$ | | $F_4 = f_0, f_1, \neg f_2, f_3, f_4, f_5, \neg f_6, f_7, \neg f_8, f_9, f_{10}$ |

The same is true for contexts since they are dependent on the relationships require and exclude, as well as, the specification of CCFs. Such relationships and CCFs compose the adaptation rules that affect the activation and deactivation of the system's features. In the running example, the ILP solver suggested different configurations when we changed the prioritization of contexts. For example, we observed different optimal configurations in the scenarios with context's prioritization $c_2 > c_6$ and $c_6 > c_2$. In the latter, for CCF as $ccf_1$, the optimal configuration is $F_4$. For the former case, the suggested configuration is $F_3$.

Therefore, the approach can support software engineering in the trade-off simulation in order to identify the impacts of prioritization changes. Notice that these changes can be enacted at runtime and, thus, it is important to know the possible effects of a priority.

### 5.3 Analyzing changes in the context feature

We also designed simulations related to changes in CCFs. Each CCF must be based on the relationship between **context feature** and its respective **context group**. In the running example, we selected only one **context feature** from each **context group** due to their XOR-group relationship. As a result, we obtained a total of six CCFs, *i.e.*, $3*2$ possible context feature changes ($ccf_1$–$ccf_6$), as shown in Table 9. Therefore, we simulated six scenarios, which were based on the different CCFs, and performed the configuration process in order to find optimal configurations that meet all constraints. For this simulation, all scenarios correspond to the same prioritization of contexts, goals, and soft goals previously presented (contexts: $c_2 > c_6$; goals: $g_2 > g_1 > g_3$; soft goals: $sg_1 > sg_2 > sg_3$). In addition, we only consider the require and exclude relationships between **context features** and system' features for a certain CCF.

The ILP solver suggested, for instance, that optimal configuration $F_1$ (see Table 10) meets $ccf_1$ (see Figure 6). It considered the following require and exclude relationships: *(i)* context feature **Emergency** *requires* features **Bluetooth** and **Single Node Processing**; *(ii)* context feature **High** *requires* features **Wifi** and **Distributed Processing**; *(iii)* context feature **Emergency** *excludes* features **Wifi** and **Distributed Processing**; and *(iv)* context feature **High** *excludes* features **Bluetooth** and **Single Node Processing**. This reasoning is used for all CCFs in order to find the optimal configurations that satisfy them.

Table 8. Corresponding $ccf$ for each optimal configuration suggested by solver ILP during the simulations.

| | | |
|---|---|---|
| | **Change of Prioritization for Soft goals** | |
| $P_1$ | $ccf_1 = <c_3,c_7>$ | $F_3$ |
| | $ccf_2 = <c_4,c_7>$ | $F_4$ |
| | $ccf_3 = <c_5,c_7>$ | $F_4$ |
| | $ccf_4 = <c_3,c_8>$ | $F_4$ |
| | $ccf_5 = <c_4,c_8>$ | $F_1$ |
| | $ccf_6 = <c_5,c_8>$ | $F_3$ |
| $P_2$ | $ccf_1 = <c_3,c_7>$ | $F_1$ |
| | $ccf_2 = <c_4,c_7>$ | $F_3$ |
| | $ccf_3 = <c_5,c_7>$ | $F_1$ |
| | $ccf_4 = <c_3,c_8>$ | $F_2$ |
| | $ccf_5 = <c_4,c_8>$ | $F_1$ |
| | $ccf_6 = <c_5,c_8>$ | $F_1$ |
| | **Change of Prioritization for Goals** | |
| $P_3$ | $ccf_1 = <c_3,c_7>$ | $F_4$ |
| | $ccf_2 = <c_4,c_7>$ | $F_4$ |
| | $ccf_3 = <c_5,c_7>$ | $F_2$ |
| | $ccf_4 = <c_3,c_8>$ | $F_4$ |
| | $ccf_5 = <c_4,c_8>$ | $F_1$ |
| | $ccf_6 = <c_5,c_8>$ | $F_3$ |
| $P_4$ | $ccf_1 = <c_3,c_7>$ | $F_4$ |
| | $ccf_2 = <c_4,c_7>$ | $F_4$ |
| | $ccf_3 = <c_5,c_7>$ | $F_2$ |
| | $ccf_4 = <c_3,c_8>$ | $F_4$ |
| | $ccf_5 = <c_4,c_8>$ | $F_1$ |
| | $ccf_6 = <c_5,c_8>$ | $F_3$ |
| | **Change of Prioritization for Contexts** | |
| $P_5$ | $ccf_1 = <c_3,c_7>$ | $F_3$ |
| | $ccf_2 = <c_4,c_7>$ | $F_4$ |
| | $ccf_3 = <c_5,c_7>$ | $F_4$ |
| | $ccf_4 = <c_3,c_8>$ | $F_4$ |
| | $ccf_5 = <c_4,c_8>$ | $F_3$ |
| | $ccf_6 = <c_5,c_8>$ | $F_3$ |
| $P_6$ | $ccf_1 = <c_3,c_7>$ | $F_4$ |
| | $ccf_2 = <c_4,c_7>$ | $F_4$ |
| | $ccf_3 = <c_5,c_7>$ | $F_3$ |
| | $ccf_4 = <c_3,c_8>$ | $F_4$ |
| | $ccf_5 = <c_4,c_8>$ | $F_1$ |
| | $ccf_6 = <c_5,c_8>$ | $F_3$ |

Likewise, the software engineer must only consider the relationships between hard goals and soft goals that correspond with a certain CCF. Figure 7 shows relationships that were considered, as follows: *(i)* hard goal **Bluetooth** with soft goals **Energy Efficiency** and **FaultTolerance**; *(ii)* hard goal **Wifi** with soft goals **Energy Efficiency** and **FaultTolerance**; *(iii)* hard goal **Single node processing** with soft goals **Energy Efficiency** and **Prediction Accurancy**; *(iv)* hard goal **Distributed processing** with soft goals **Energy Efficiency** and **Prediction Accurancy**;

Table 9. Its presents the Require and exclude relationships for the valid $ccfs$

| CCF | Relationship between context feature and system' feature | | Optimal Configuration |
|---|---|---|---|
| | Require | Exclude | |
| $ccf_1 =$ $< c_3,c_7 >$ | $< c_3, f_2 >$ $< c_3, f_8 >$ $< c_7, f_3 >$ $< c_7, f_9 >$ | $< c_3, f_3 >$ $< c_3, f_9 >$ $< c_7, f_2 >$ $< c_7, f_8 >$ | $F_1$ |
| $ccf_2 =$ $< c_4,c_7 >$ | $< c_4, f_9 >$ $< c_7, f_3 >$ $< c_7, f_9 >$ | $< c_4, f_8 >$ $< c_7, f_2 >$ $< c_7, f_8 >$ | $F_1$ |
| $ccf_3 =$ $< c_5,c_7 >$ | $< c_5, f_3 >$ $< c_5, f_5 >$ $< c_7, f_3 >$ $< c_7, f_9 >$ | $< c_5, f_2 >$ $< c_5, f_6 >$ $< c_7, f_2 >$ $< c_7, f_8 >$ | $F_2$ |
| $ccf_4 =$ $< c_3,c_8 >$ | $< c_3, f_2 >$ $< c_3, f_8 >$ $< c_8, f_5 >$ $< c_8, f_8 >$ | $< c_3, f_3 >$ $< c_3, f_9 >$ $< c_8, f_3 >$ $< c_8, f_9 >$ | $F_1$ |
| $ccf_5 =$ $< c_4,c_8 >$ | $< c_4, f_9 >$ $< c_8, f_5 >$ $< c_8, f_8 >$ | $< c_4, f_8 >$ $< c_8, f_3 >$ $< c_8, f_9 >$ | $F_3$ |
| $ccf_6 =$ $< c_5,c_8 >$ | $< c_5, f_3 >$ $< c_5, f_5 >$ $< c_8, f_5 >$ $< c_8, f_8 >$ | $< c_5, f_2 >$ $< c_5, f_6 >$ $< c_8, f_3 >$ $< c_8, f_9 >$ | $F_1$ |

Table 10. Optimal configurations suggested by solver ILP

| Configuration | System' features |
|---|---|
| $F_1$ | $f_0, f_1, f_2, \neg f_3, f_4, f_5, \neg f_6, f_7, f_8, \neg f_9, f_{10}$ |
| $F_2$ | $f_0, f_1, f_2, \neg f_3, f_4, f_5, \neg f_6, f_7, \neg f_8, f_9, f_{10}$ |
| $F_3$ | $f_0, f_1, \neg f_2, f_3, f_4, f_5, \neg f_6, f_7, f_8, \neg f_9, f_{10}$ |

Table 11. Utility values of the system's features

| | Features | | | | | |
|---|---|---|---|---|---|---|
| Configuration | $f_2$ | $f_3$ | $f_5$ | $f_6$ | $f_8$ | $f_9$ |
| $F_1$ for $ccf_1$ | 1.169 | 0.277 | 0 | 0 | 1.080 | -0.043 |
| $F_2$ for $ccf_3$ | 0.003 | 0.849 | 1.130 | -0.130 | 0.496 | 0.372 |
| $F_3$ for $ccf_5$ | 1.336 | -0.150 | 0 | 0 | 0.746 | 0.289 |

In this analysis, the ILP solver suggested three different optimal configurations ($F_1 - F_3$), as shown in Table 10. Each optimal configuration, derived from *GridStix* DSPL, is composed of a set of system features. Table 11 presents the system's features with their respective utility values. Those three optimal configurations correspond to the CCFs $ccf_1$, $ccf_3$, and $ccf_5$, respectively.

Fig. 6. An example of CCF for Grid Stix DAS: $ccf_1$ (Emergency and High)

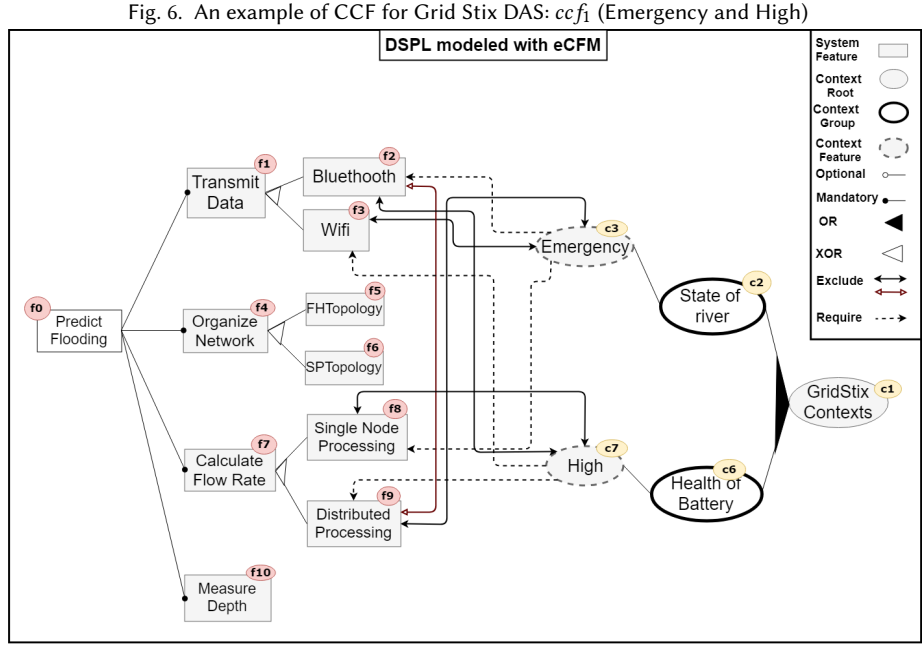

For the CCF $ccf_1$, the ILP solver suggested $F_1$ as the optimal configuration. In this configuration, features $f_2$, $f_5$, and $f_8$ were selected. By observing the utility values, they indicate that features $f_3$ and $f_9$ have the lowest values $\mathbb{C}(f_i)$. However, although the features $f_5$ and $f_6$ have presented equal values $\mathbb{C}(f_i)$, the solver suggested that feature $f_5$ should be selected. It reveals that the solver suggests the first one among all features with equal values. The same occurred when configuration $F_3$ was considered optimal for the CCF $ccf_5$, which corresponds to the activation of features $f_2$, $f_5$, and $f_8$. For the CCF $ccf_3$, the configuration $F_2$ was suggested as optimal. In this case, features $f_3$, $f_5$, and $f_8$ were selected since they presented greater utility values than feature $f_2$, $f_6$, and $f_9$.

### 5.4 Definition of adaptation model

From a specific CCF, it is possible to define dynamic adaptation models. These models show how the DSPL application can evolve from one CCF to another changing its respective optimal configuration. Figure 8 shows an example of a DSPL adaptation model composed of three optimal configurations, which can be loaded by six different CCFs ($ccf_1$–$ccf_6$).

In this example, we chose arbitrarily the configuration $F_1$ as initial due to its recurrence in the majority of simulations. However, the software engineer can choose another one. It is possible to implement eight adaptations by considering the CCF changes. They can be detected in the runtime environment, as follows: *(i)* when the CCFs $ccf_1$, $ccf_2$, $ccf_4$, and $ccf_6$ are detected, the configuration $F_1$ is loaded; *(ii)* the configuration $F_2$ is loaded when CCF $ccf_3$ is detected; and *(iii)* when CCF $ccf_5$ is detected, configuration $F_3$ is loaded. Therefore, this process should be made for all CCF changes and their respective optimal configurations.

Each adaptation model encompasses a specific prioritization of *goals*, *soft goals*, and *contexts*. Then, an adaptation model can be considered a product in DSPL engineering. In addition, the variability decisions are made at runtime by meeting the CCF changes. Thus, it provides support for the entire range of adaptations handled during application engineering.

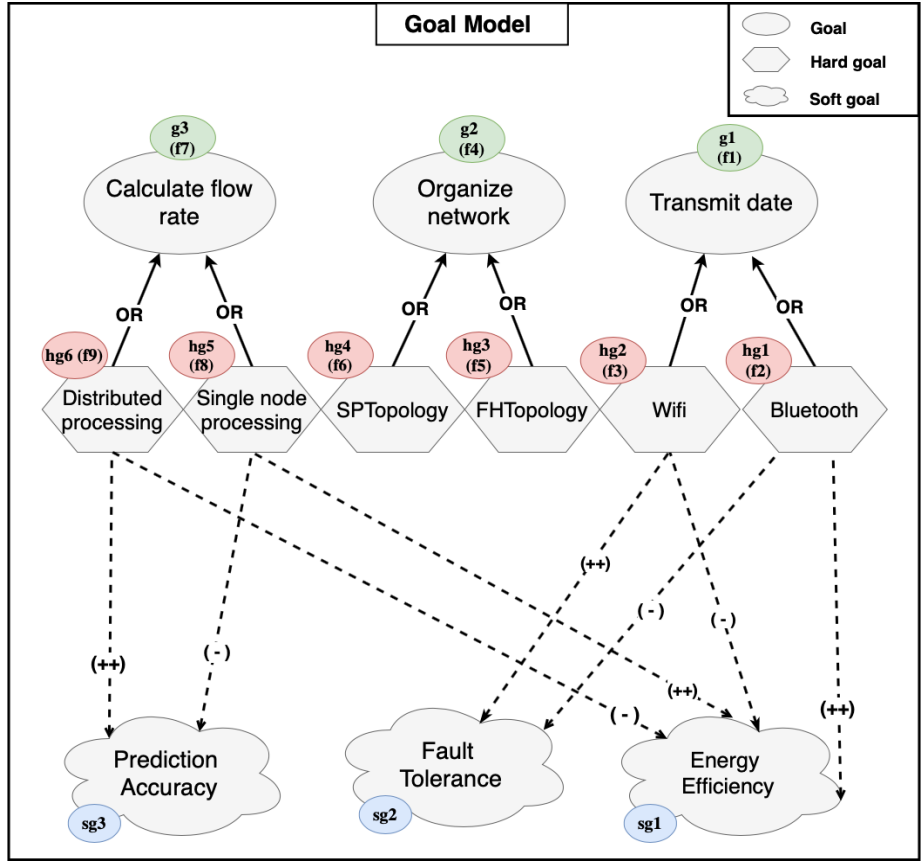

Fig. 7. Goal model corresponding to CCF $ccf_1$ (Emergency and High)

Indeed, adaptation models offer the potential to facilitate communication between software engineers and developers. Bencomo *et al.* [8] proposed the development of similar adaptation models to represent the transitions among contextual changes. These assets facilitate the development of DSPL applications and the understanding of how they can behave from a certain context change.

## 6 LESSONS LEARNED

Variability management is an important activity that describes different configurations of the system. This activity requires a consistent and scalable approach to explore, define, represent, implement, and evolve DSPL. Based on simulations, we evidenced that our approach can be used for such purpose, *i.e.* it aims to explore reuse and support the specification of adaptation models for both dimensions *structural variability* and *context variability*.

We performed simulations with the DSPL *GridStix*, as presented in Section 5. Such simulations aimed to verify the feasibility of using the ToffA-DSPL approach and encourage the developers to use it in the configuration selection process of DSPL. It also was possible to identify how to develop a generic optimization model, which can be used for different domains and system applications.

Fig. 8. An example of an adaptation model for a DAS application. It is defined after context feature change analysis. Each configuration represents a set of features suggested as valid and optimal during the configuration selection process.

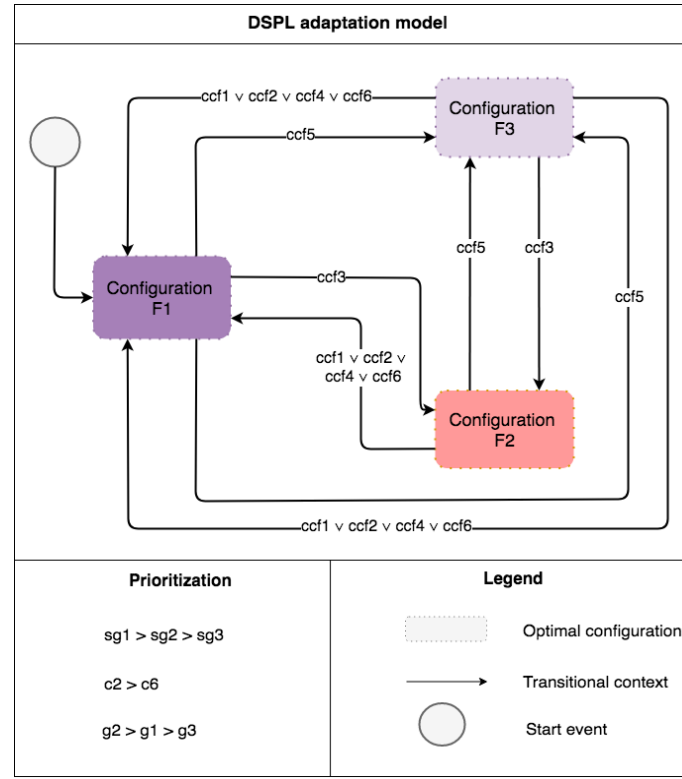

As a result of the simulations, we concluded that the ToffA-DSPL approach is useful for performing trade-off analysis, generating optimal configurations, and identifying possible adaptations that can occur at runtime. It meets the *structural variability* and *context variability*, besides the different measurements of *prioritization*, *contribution*, and *satisfaction levels* assigned to *goals*, *soft goals*, and *contexts*. However, the way of building the eCFM and goal model may impact heavily on the result suggested by the ILP solver. Considering this factor, we provide the following points for consideration:

(i) Inconsistency in CCFs may arise when the dependencies between features and **context features** are not represented correctly. The same is true with dependencies between goals and hard goals. In this scenario, the software engineer and stakeholders should check the models aiming to identify design faults. After that, they can agree on developing DAS applications according to the specialized feature model.

(ii) During the simulations, the optimization model identified some design faults. Initially, for example, we inserted require and exclude relationships between parent features and **context features**. At the same time, we inserted them between the leaf features of each control system and **context features**. It resulted in faults, which we corrected to continue the simulations. Therefore, we recommend software engineers avoid modeling and developing DAS applications with such adaptation rules to prevent failures at runtime.
(iii) A problem associated with to use of a utility function as an optimization strategy is the difficulty of defining such a function that precisely represents the stakeholder's preferences. It aims to the heuristic representation of a desirable configuration. However, to find such a configuration, it is necessary to measure a utility value for each system's feature, which equals the weighted sum of all values assigned to the modeling elements. Then, the solver suggests an optimal configuration among possible configurations that maximize this utility value.
(iv) An adaptation to a given CCF corresponds to a product's optimal configuration. Therefore, during application engineering, the DAS applications should be built according to variations in the requirement prioritization and artifacts defined in domain engineering. Next, CCFs must be predetermined for all possible dynamic adaptations, to define different adaptation models. Thus, the software engineer can choose one of them to be eventually developed.

## 7 LIMITATIONS

In this section, we provide an extra discussion on the limitations of ToffA-DSPL, which were identified after carrying out the exploratory study.

**Trade-off analysis at runtime.** Our approach aims to identify at design time a set of possible adaptations and information that can affect the product configuration. However, the system quality evaluation must also be made at runtime to check the capacity of the system to meet self-adaptive operations [17]. Thus, we also intend to perform an investigation of the feasibility of using the approach for trade-off analysis during the execution of the DAS application. It aims to conduct adaptations at runtime by meeting the desirable variants and default configurations defined in the adaptation model.

**Technical aspects.** The definition of the degree to which the variable features satisfy the *soft goals* is often subjective and makes the configuration selection process more difficult. It depends on the software engineer's knowledge of the application domain and may not be optimal to satisfy certain change requests at runtime [80]. Further efforts should be made, regarding the employ of the ToffA-DSPL approach at the running application to establish confidence regarding assigning suitable satisfaction levels for NFRs. Therefore, more studies are needed in distinct domains considering, especially, the point of view of the DAS developers.

**Uncertainty.** Although we employ the trade-off analysis to find a set of valid and optimal configurations that meet different scenarios, the DAS applications are subjected to uncertainty that was not explicitly designed. Such uncertainty is caused by parameters whose values change at the running applications and can make NFRs unsatisfied. The parameters are given information related to sensor failure or occlusion which can occur at runtime [11, 26]. Importantly, we may propose to evolve the ToffA-DSPL approach aiming to address uncertainty and define a strategy to handle the possible adaptations that were not foreseen at design time.

**Tool support.** In our current approach only STEP 5 (optimization) is automated. However, it is not an easy task to measure the satisfaction level, prioritization, and contribution to a huge number of model elements such as features, contexts, and NFRs. For systems that present many modeling elements and to overcome scalability issues, as future work, we intend to develop a tool that encompasses all steps of the ToffA-DSPL approach.

## 8 CONCLUDING REMARKS

We developed the ToffA-DSPL approach aiming to identify feasible configurations. Concerning identifying feasible configurations, ToffA-DSPL deals with the configuration selection process embracing the interactions between contexts and NFRs. Such an approach uses the utility function as a strategy to express the priorities of users over services provided by DSPL applications. Those priorities are represented as weights aiming to direct the choice of a feasible solution.

We performed a study based on simulations using the *GridStix* DAS (Section 5) to gather initial evidence about the feasibility of using the ToffA-DSPL approach from the point of view of *(i)* conduction of trade-off analysis by considering changes in the prioritization of elements that comprise eCFM and goal models such as goals, soft goals, and contexts; and *(ii)* definition of adaptation models, from optimal configurations found in the CCF change analysis. All simulations presented consistent results following the real-world scenarios and satisfied the estimated utility values and linear constraints. They met the variability dimensions and different satisfaction levels assigned to *soft goals*, besides measurements of prioritization assigned to *contexts*, *goals*, and *soft goals*.

In future work, we plan to perform empirical studies to improve the approach proposed in this paper. In addition, we plan to investigate the application of other optimization methods for dealing with configuration selection analysis. Although the assessment results have shown that using ILP is quite efficient for identifying optimal configurations, studies existing [19, 58] in the literature reported evaluations when multi-objective algorithms such as NSGA-II, IBEA, and PAES were compared with each other by demonstrating good results. Thus, we plan to compare different multi-objective algorithms with diverse configurations to identify the best option to be used with our approach and benefit the design and development of DAS by using the DSPL engineering processes.